\documentclass[proof]{pasj01}
\usepackage{natbib}
\usepackage{color}
\usepackage{bm}
\usepackage{subfigure}
\newcommand{\gas}{\rm gas}

\newcommand{\HAGN}{H$_{\rm AGN}$}
\newcommand{\HnoAGN}{H$_{\rm noAGN}$}
\newcommand{\HDM}{H$_{\rm DM}$}

\begin{document}
\title{
Projected Axis Ratios of Galaxy Clusters in the Horizon-AGN Simulation:
Impact of Baryon Physics and Comparison with Observations}
\author{
Daichi \textsc{Suto} \altaffilmark{1}, 
S\'ebastien \textsc{Peirani} \altaffilmark{2,3}, 
Yohan \textsc{Dubois} \altaffilmark{2},
Tetsu \textsc{Kitayama} \altaffilmark{4}, 
Takahiro \textsc{Nishimichi} \altaffilmark{3,5},
Shin \textsc{Sasaki} \altaffilmark{6},
and Yasushi \textsc{Suto} \altaffilmark{1,7}
}
\email{suto@phys.s.u-tokyo.ac.jp}
\altaffiltext{1}{Department of Physics, The University of Tokyo, Tokyo
113-0033, Japan}
\altaffiltext{2}{CNRS and UPMC Universit\'e Paris 06, UMR 7095, Institut
d'Astrophysique de Paris, 98 bis Boulevard Arago, Paris 75014, France}
\altaffiltext{3}{Kavli Institute for the Physics and Mathematics of the
Universe (WPI), The University of Tokyo Institutes for Advanced Study,
The University of Tokyo, 5-1-5 Kashiwanoha, Kashiwa 277-8583, Japan}
\altaffiltext{4}{Department of Physics, Toho University, Funabashi,
Chiba 274-8510, Japan}
\altaffiltext{5}{CREST, JST, 4-1-8 Honcho, Kawaguchi, Saitama, 332-0012,
Japan}
\altaffiltext{6}{Department of Physics, Tokyo Metropolitan University,
Hachioji, Tokyo 192-0397, Japan}
\altaffiltext{7}{Research Center for the Early Universe, School of
Science, The University of Tokyo, Tokyo 113-0033, Japan}

\maketitle

\begin{abstract}
We characterize the non-sphericity of galaxy clusters by the projected
axis ratio of spatial distribution of star, dark matter, and X-ray
surface brightness (XSB). We select 40 simulated groups and clusters of
galaxies with mass larger than $5\times10^{13} M_\odot$ from the Horizon
simulation that fully incorporates the relevant baryon physics, in
particular, the AGN feedback. We find that the baryonic physics around
the central region of galaxy clusters significantly affects the
non-sphericity of dark matter distribution even beyond the central
region, approximately up to the half of the virial radius.
{Therefore it is very difficult to predict the the probability
density function (PDF) of the projected axis ratio of XSB from}
{dark-matter only} {N-body simulations as attempted in
previous} {studies}.  Indeed we find that the PDF derived from our
simulated clusters exhibits much better agreement with {that from
the observed X-ray clusters}.  {This indicates that our present
methodology to estimate the non-sphericity directly from the Horizon
simulation is} {useful and} {promising.  Further improvements
in both numerical} {modeling} {and observational data will
establish the non-sphericity of clusters} {as} {a cosmological
test} complementary to more conventional statistics based on spherically
averaged quantities.
\end{abstract}

\section{Introduction}

Statistics of galaxy clusters, along with cosmic microwave background
and large-scale distribution of galaxies, has made significant
contribution to {the establishment of} the standard cosmological model dominated
by cold dark matter (CDM). Indeed the resulting CDM paradigm has passed
a variety of theoretical and observational tests, except for somewhat
controversial problems on small scales.  {While there are a number
of important successes in terms of the spherically averaged} {properties}
{\citep[e.g.,][]{Press74,Navarro96,Navarro97,Suto16a}, observed galaxy
clusters are} {typically} {far from spherical.}  Nevertheless a majority of
previous studies on galaxy clusters has not properly taken into account
their non-sphericity.

In the previous paper \citep{Suto16b}, we found that the evolution of
non-sphericity in N-body simulations is not in quantitative agreement
with a simple model of ellipsoidal collapse. We also found that the
probability density function (PDF) of projected axis ratios of dark
matter halos is empirically well approximated by {the beta function},
which is fairly insensitive to the mass and redshift of those halos.
While the latter result is potentially useful in confronting the result
with the data of weak lensing halos, there are a couple of {fundamental
limitations. One is that} the currently available number of
{good-quality weak-lensing} data is quite limited. {The other is
that} the axis ratio from the lensing shear map needs to be measured
without assuming the self-similarity in the non-sphericity of halos
{\citep[see, {\it e.g.,}][]{Oguri10,Suto16b}}.

Therefore we consider, instead, the non-sphericity of the X-ray surface
brightness {(XSB)} $S_X$ of galaxy clusters in the present paper.
Since $S_X$ is one of the primary observables for galaxy clusters, a
large number of high-quality data-sets {are} already available, and also is
expected to further increase in the near future.  In fact,
\cite{Kawahara10} has measured the non-sphericity of the XSB of {61}
galaxy clusters, and reported that the PDF of their axis ratio is
consistent with the prediction on the basis of \citet{Jing02} and
\citet{Oguri03}.

{We will revisit the problem in the present paper because} the
theoretical prediction of non-sphericity of $S_X$ is more difficult than
the observational measurements due to the complicated effects of baryons
in galaxy clusters \citep{Debattista08,Teyssier11,Bryan13,Butsky15,Cui16}.

The distribution of gas is {generally} different from that of dark matter
{in clusters} \citep{Lee03}. Under the assumption of hydrostatic equilibrium (HSE),
the gas mass density $\rho_{\gas}$ satisfies the following equation:
\begin{equation}
\label{eq:hse}
\frac{1}{\rho_{\gas}}\nabla p=-\nabla\phi,
\end{equation}
where $p$ is the gas pressure, and $\phi$ is the gravitational potential
mainly determined by the dark matter density distribution.  If the gas
is isothermal, Equation (\ref{eq:hse}) reduces to $\nabla \log
\rho_{\gas} \propto \nabla \phi$. Thus the resulting gas distribution
traces the {\it isopotential} surface, instead of {the} {\it isodensity}
surface {of the underlying dark matter}. Since the isopotential 
surfaces tend to be rounder than the isodensity surfaces, 
gas distribution becomes more spherical than that
of dark matter \citep{Lee03}.

In reality, the conventional assumption of HSE is not so accurate
\citep{Lau09,Lau13,Fang09,Suto13} due to the dynamical motion of gas.
For those reasons, the non-sphericity of gas distribution {cannot}
be related to that of dark matter {distribution in a straightforward
fashion}.  Hence it is essential to use numerical simulations including
gas physics in order to precisely study the non-sphericity of gas
density, and therefore that of $S_X$ of galaxy clusters. This is exactly
what we will {address} in this paper.

The rest of the paper is organized as follows.  Section
\ref{sec:numerical} briefly describes the Horizon simulation that we use
in modeling the non-sphericity of galaxy clusters.  We emphasize the
important role of baryon physics, in particular, the active galactic
nuclei {(AGNs)} feedback by considering spherically-averaged density and
temperature profiles.  Then we measure the axis ratios of {dark
matter, stellar, and XSB distributions for simulated galaxy clusters} in
Section \ref{sec:statistics}.  Section \ref{sec:comparison} compares the
PDF of the axis ratios constructed from 40 simulated clusters against
that computed by \citet{Kawahara10} {for observed 61 clusters}. Final section
is devoted to summary and discussion.

\section{Numerical Modeling \label{sec:numerical}}

\subsection{Horizon simulation \label{subsec:horizonsimulation}}

Our current study is entirely based on samples of simulated groups and
clusters of galaxies extracted from three cosmological hydrodynamical
{runs}, Horizon-AGN (\HAGN), Horizon-noAGN (\HnoAGN), and Horizon-DM
(\HDM).  Results of \HAGN~ are already described in detail in
\citet{Dubois14}, so we {briefly describe here the major feature of
those runs that is} relevant to our current discussion.

The simulation adopts a set of cosmological parameters derived from the
Wilkinson Microwave Anisotropy Probe 7 year \citep{Komatsu11};
$\Omega_{m,0}=0.272$, $\Omega_{\Lambda,0}=0.728$, $\Omega_{b,0}= 0.045$,
$\sigma_8=0.81$, $H_0=70.4$ km s$^{-1}$ Mpc$^{-1}$, and $n_s=0.967$.
The simulation runs employ $1024^3$ dark matter particles in a periodic
cube with a side length of 100 $h^{-1}$ Mpc, which results in a dark
matter mass resolution of $8.27\times10^7 M_\odot$.  {The three runs
adopt the identical initial conditions that are generated with the {\tt
MPGRAFIC} software \citep{Prunet08}, except the fact that the dark
matter density parameter in \HDM \, is set as $\Omega_{m,0}$, instead of
$\Omega_{m,0}-\Omega_{b,0}$ in \HAGN \, and \HnoAGN.}

Hydrodynamics of gas and other baryon physics {is} solved on grids over
the simulation box. The size of gas cells is initially set to 136 kpc,
and is refined subsequently when the number of dark matter particles in
a cell becomes more than eight, or when the total baryonic mass in a
cell {becomes} eight times the dark matter mass resolution
($8.27\times10^7 M_\odot$). The refinement is carried out up to 7 times,
and therefore the minimum cell size is 1.06 kpc.

The radiative cooling of gas due to H, He, and metals is {modeled}
according to \cite{Sutherland93}. Heating from a uniform UV background
is also implemented following \cite{Haardt96}. Star particles are
created according to the Schmidt-Kennicutt law using a random Poisson
process \citep{Rasera06, Dubois08} if the gas hydrogen number density in
a cell exceeds the threshold of $n_0=0.1$ cm$^{-3}$.  In addition,
feedback from stellar winds, supernovae {(SNe)} type Ia and II are
also taken into account for mas, energy, and metal release.

{Type II SNe are taken into account assuming the Salpeter initial
mass function. The} {SN} {energy is released into the surrounding gas
according to the Sedov blast wave solution because the thermal input is
radiated away due to the efficient gas cooling in high-density regions.
The frequency of Type Ia SN explosions follows \cite{Greggio83}, and the
mechanical energy from Type II SNe is taken from {\tt STARBURST99}
\citep{Leitherer99,Leitherer10}. }

It is known that the feedback from {AGNs} plays a
significant role in the evolution of a central part of galaxy clusters.
The AGN feedback in the present simulations is incorporated following
\cite{Dubois12}. Black holes {(BHs)} of an initial seed mass $10^5
M_\odot$ are created when the gas mass density $\rho_{\rm gas}$ in a
cell exceeds $m_{\rm H} n_0$, {with $m_{\rm H}$ being the hydrogen
mass}. Since the accretion onto BHs cannot be resolved in the current
simulations, their growth is {empirically computed adopting} the
Bondi-Hoyle-Lyttleton accretion rate $\dot{M}_{\rm BH}$.  Depending on
the ratio $\chi \equiv \dot{M}_{\rm BH}/\dot{M}_{\rm Edd}$ relative to
the Eddington rate, the AGN feedback is divided into two different
modes.  In the radio mode ($\chi<0.01$), the feedback energy is ejected
into a bipolar outflow with a jet velocity of $10^4$ km s$^{-1}$
following \cite{Omma04}.  Otherwise, the feedback is in the quasar mode
($\chi>0.01$) in which thermal energy is isotropically injected into
gas.

The first {simulation}, \HAGN, is a full hydrodynamic run with gas
cooling and both SN and AGN feedback. The second {one}, \HnoAGN, is a
hydrodynamic simulation with gas cooling and SN feedback, but without
AGN feedback. Finally \HDM \, is a dark matter only {run} without
baryon physics.  The first is our main {simulation}, and the other two are used
{for reference} to examine the effect of AGN feedback and baryon physics.
The three runs {(\HAGN\,, \HnoAGN\,, and \HDM)} \, are performed
using the the same initial conditions and sub-grid modeling.

We extract {all} 40 halos with mass larger than $5\times10^{13}
M_\odot$ from \HAGN\, using the AdaptaHOP halofinder \citep{Aubert2004}.
{Since those halos are not spherical, we decided to characterize them in
terms of $M_{200}$, the mass of a sphere within which the mean density
is 200 times the cosmic critical density at $z=0$. In most cases, the
resulting value of $M_{200}$ is close to that defined by the AdaptaHOP
halofinder, but in some cases is smaller because of the presence of
substructures.}

In reality, the mass scale of the halos corresponds to that of groups
and clusters of galaxies, but we simply call them clusters in what
follows.  {The counterparts of the clusters in three different runs
are identified as follows.  Since we start from the same initial
conditions, each dark matter particle} {shares} {a common
identity between the 3 simulations.} {If more than 75\% of the}
{member} {particles of a given halo from \HAGN\, are found in
another halo from \HnoAGN\, and \HDM\, runs, then we assume that}
{these}\ red{objects are the counterparts of \HAGN\, run.}
{Indeed, we can find the counterparts for all the 40 clusters with
this procedure.}

  In the following sections, we analyze the non-sphericity of
{XSB} for the two sets of 40 clusters in
\HAGN~ and \HnoAGN. For comparison, {we also discuss} the non-sphericity 
of dark matter halos for the three sets of 40 clusters.

{Of course, no simulation is perfect, and \HAGN\, should be rather
regarded as one of the most successful simulation runs.  As we will show
below, the non-sphericity of galaxy clusters is very sensitive to the
baryon physics, and we have to keep in mind that \HAGN\, is nothing more
than an empirically calibrated model at this point. Nevertheless, 
\HAGN\, proves to be a very useful model that avoids
several {\it ad-hoc} assumptions adopted in {previous studies}, and
significantly improves the predictions of the non-sphericity of galaxy
clusters.  Also the quantitative comparison with \HnoAGN\, and \HDM\,
clarifies the impact of baryon physics on gas, star, and dark matter
distribution.}

\subsection{Spherically-averaged profiles of density and temperature 
\label{subsec:spherial-profile}}

Reliable modeling of {XSB} of galaxy clusters crucially depends on
baryon physics that is incorporated in the simulations. In particular,
it is widely known that AGN feedback plays an important role
\citep[e.g.,][]{Dubois10,Dubois11}. This is also clear from the
comparison between \HAGN\, and \HnoAGN.

\begin{figure}[tbh]
\begin{center}
\FigureFile(70mm,70mm){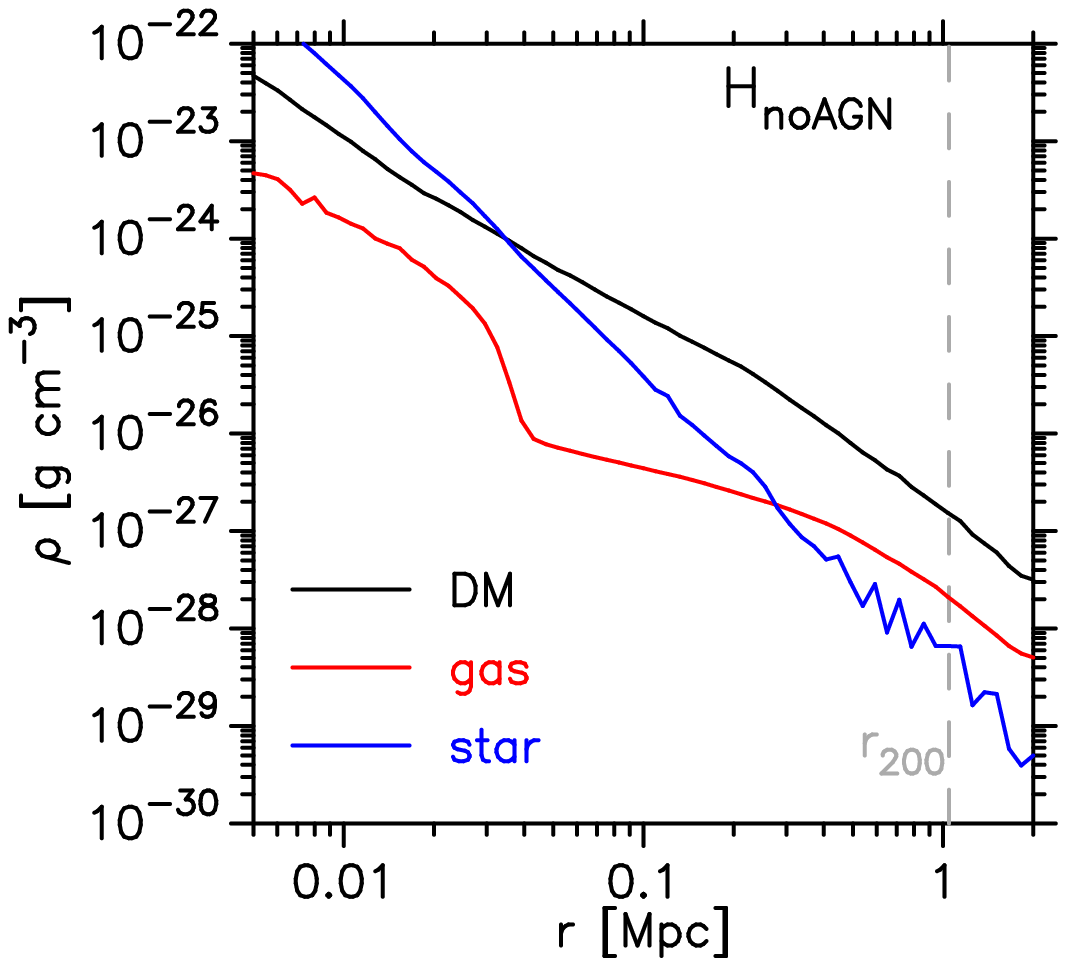}
\qquad
\FigureFile(70mm,70mm){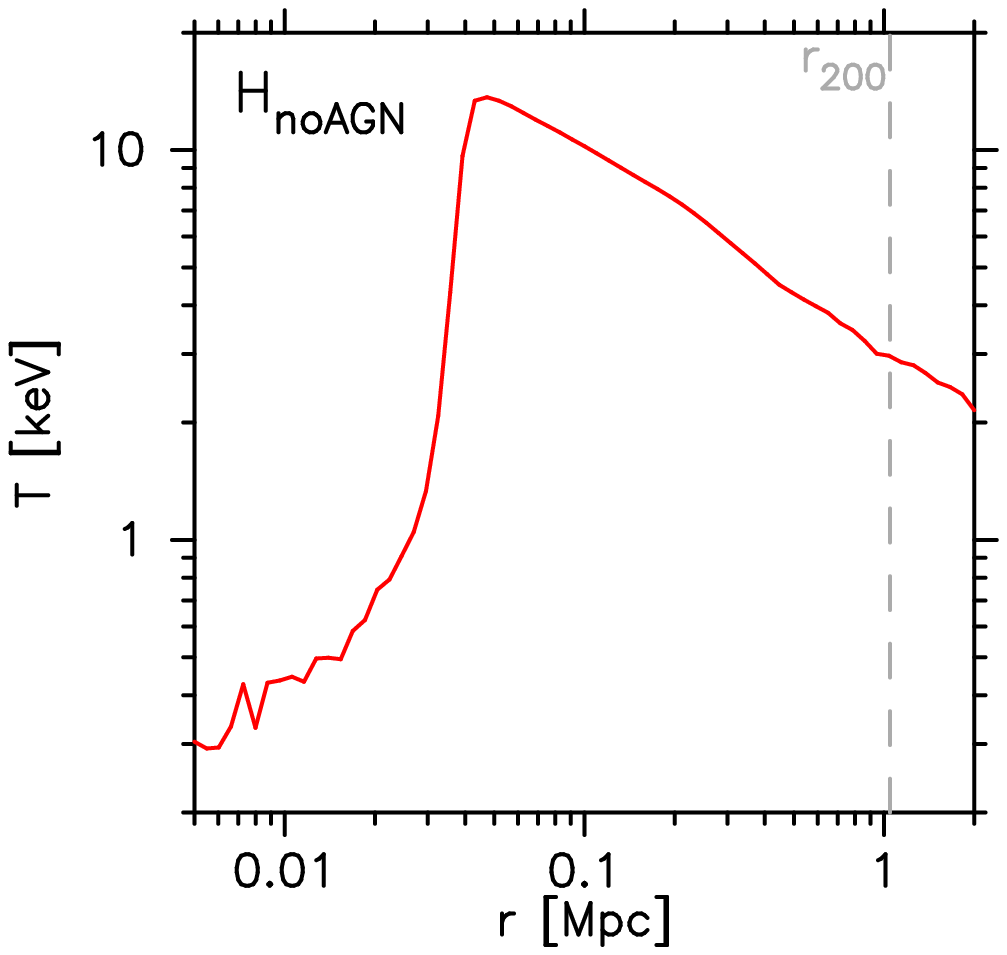}
\end{center}
\caption{Radial profiles of density (left) and mass-weighted gas
temperature (right) of the cluster with $M_{200}\sim 4.5\times 10^{14}
M_\odot$ for \HnoAGN. The density profiles of gas (red), dark matter
(black), and stars (blue) are shown in the left panel. The gray dashed
vertical line indicates $r_{200}$ of the cluster.}  \label{f5dtwo}
\begin{center}
\FigureFile(70mm,70mm){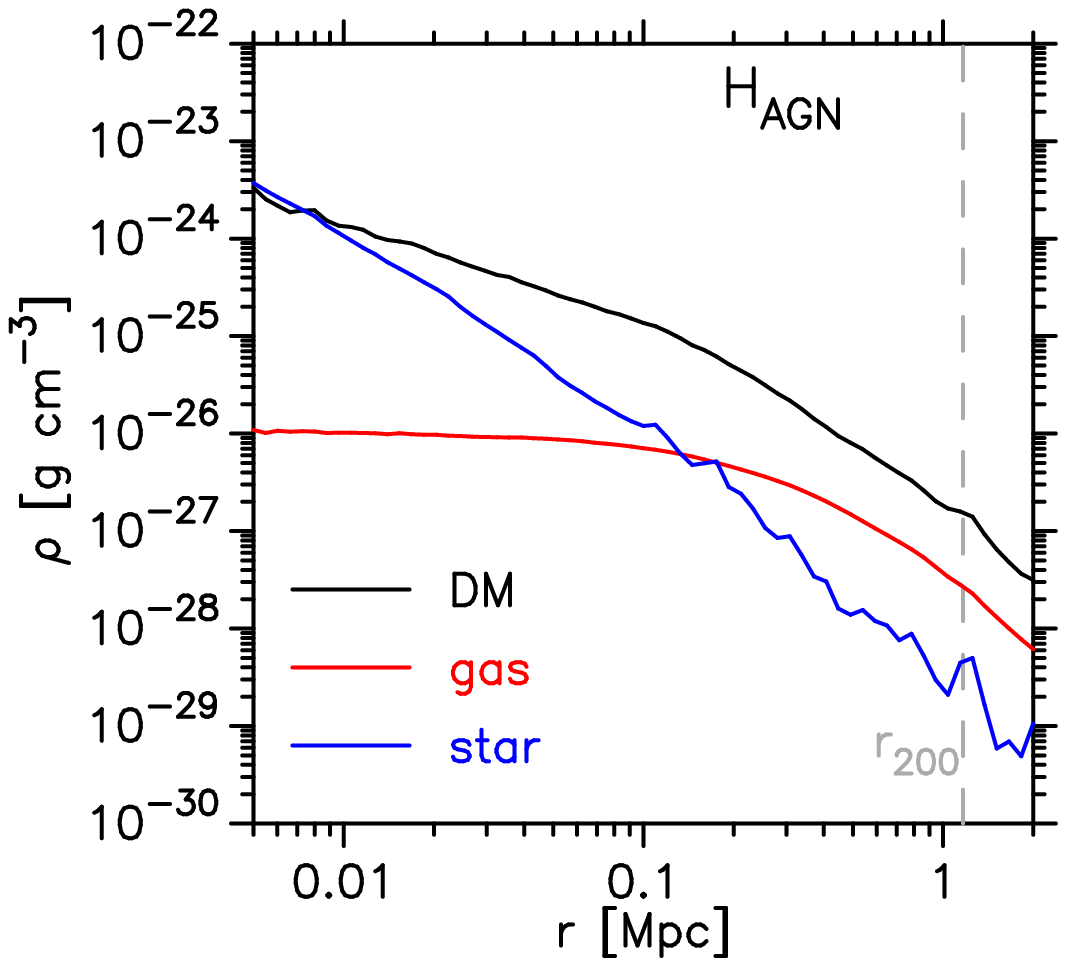}
\qquad
\FigureFile(70mm,70mm){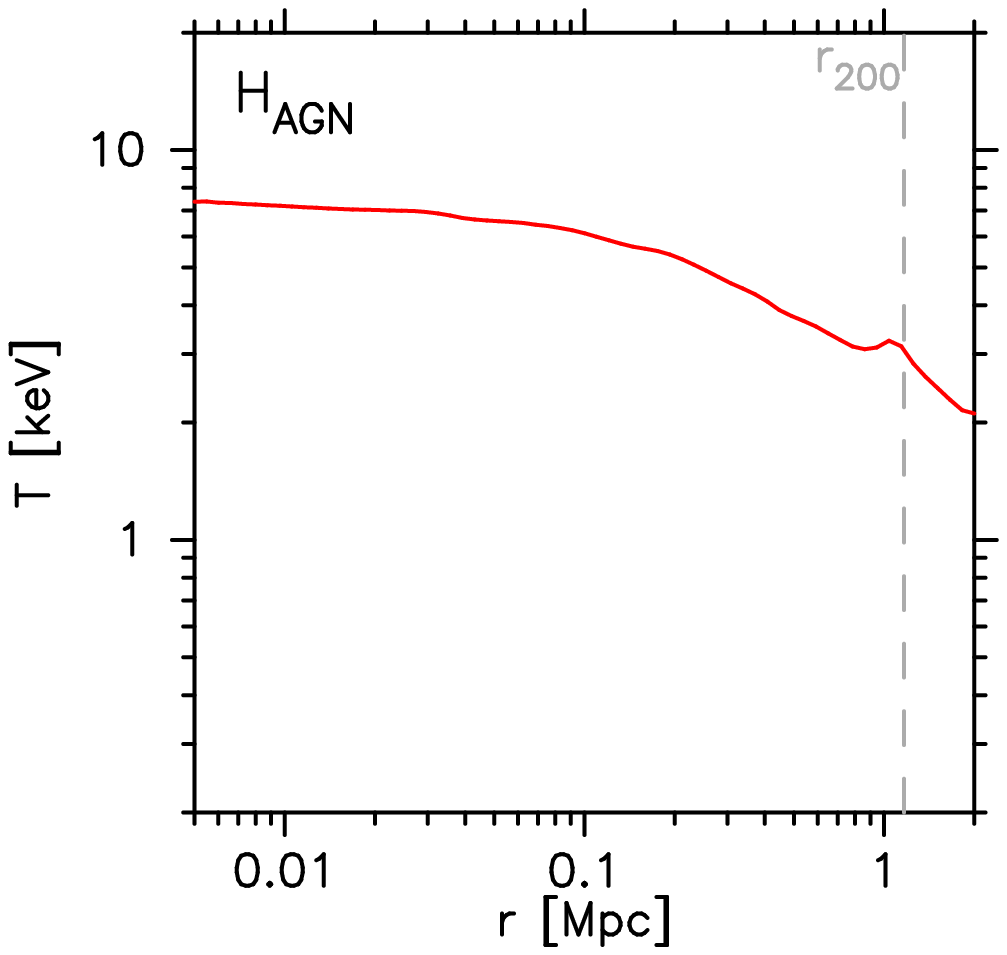}
\end{center}
\caption{Same as Figure \ref{f5dtwo}, but for the corresponding cluster
in \HAGN. Note that $r_{200}$ is slightly different between \HAGN~ and
\HnoAGN.}  \label{f5dtw}
\end{figure}

Figure \ref{f5dtwo} illustrates the density and temperature profiles of
one simulated cluster ($M_{200}\approx 4.5\times 10^{14} M_\odot$) in
\HnoAGN. In the left-panel, we plot the density profiles of gas (red),
dark matter (black), and stars (blue). Also, the (mass-weighted) gas
temperature profile is shown in the right panel. Both the strong excess
of the central stellar and gas densities and the sudden decrease of the
central gas temperature are inconsistent with the typical profiles of
observed galaxy clusters, and should be ascribed to the unrealistic
over-cooling of gas in the region. This indicates that the feedback from
SNe is not sufficient to stop star formation and gas-cooling in the
central region.

In contrast, the corresponding cluster in \HAGN\, exhibits density and
temperature profiles {that are} consistent with observed non-cool
core clusters (see Figure \ref{f5dtw}{; and also \citealt{Okabe16,Snowden08}}). 
The stellar density exceeds the
dark matter density inside $\sim$ 10 kpc, corresponding to the typical
galactic scale. The gas density profile is flat inside $\sim$ 100 kpc,
and does not exhibit a rapid increase unlike in Figure
\ref{f5dtwo}. Also, the gas temperature gradually increases toward the
center, and the temperature at around $r\lesssim 0.5 r_{200}$ is
consistent with the observed mass-temperature relation \citep{Arnaud99}.

\subsection{Estimate of non-sphericity \label{subsec:estimate}}

We use the mass tensor to estimate the non-sphericity of dark
matter density distribution:
\begin{equation}
\label{eq:masstensor}
I_{\alpha\beta}=\sum_i^N m^{(i)}x_\alpha^{(i)}x_\beta^{(i)}
\qquad
(\alpha, \beta = 1, 2),
\end{equation}
where $m^{(i)}$ and $x_\alpha^{(i)}$ are the mass and the projected
position vector of the $i$-th particle within a given enclosed mass
region (specified by the value of $N$).  The mass tensor is computed
iteratively until it is converged following \citet{Suto16b}, yielding
the semi-minor axis $a_1$ and semi-major axis $a_2$ for projected
ellipses. {In what follows}, we denote the projected axis ratio by
$q \equiv a_1/a_2 (<1)$.

{ For each of 40 simulated clusters, we calculate the projected dark
matter density distribution along the $x$-, $y$-, and $z$-axes and
determine the ellipses by using the mass tensor $I_{\alpha\beta}$.  In
later sections, we show the axis ratio $q$ of} {XSB with a fixed value
of $a_2$;} {$a_2/r_{200}=0.1$, 0.2,..., 0.9, and 1.0.  Hence we attempt
to derive $q$ from the dark matter component using the same values of
$a_2/r_{200}$, in order to make a direct comparison 
and with that from $S_X$. 
$I_{\alpha\beta}$ is, however, defined for a fixed value of the enclosed
mass, while the} {corresponding} {value of $a_2$ is not known in
advance.  We thus determine the axis ratios iteratively until the
resultant value $a_2$ is converged within one percent.  } The same
procedure is repeated for dark matter particles in \HDM, \HnoAGN,\, and
\HAGN.  The non-sphericity of stellar mass density distribution in
\HnoAGN\, and \HAGN\, is also estimated in the same fashion.

The mass tensor {defined by equation (\ref{eq:masstensor})} cannot
be directly used to estimate the non-sphericity of {XSB} 
that is defined on grids, {instead of particles}. Thus we
first compute
\begin{equation}
\label{e5sx}
S_X(\vec{\theta})=\frac{1}{4\pi (1+z)^4}\int dl\, n_{\gas}^2\Lambda(T,Z)
\end{equation}
along the line-of-sight within a sphere of $r_{200}$ {around} each
cluster center, where {$n_{\gas} = n_{\gas}(\vec{\theta},l)$} is the
number density of gas and $\Lambda(T,Z)$ is the X-ray cooling function
that depends on the gas temperature {$T(\vec{\theta},l)$} and
metallicity {$Z(\vec{\theta},l)$}.  Note that, over the typical
temperature range of galaxy clusters (1 keV $\lesssim T\lesssim$ 10
keV), {$\Lambda(T,Z)$} is approximately proportional to $T^{1/2}$, and
therefore the shape of $S_X$ is largely determined by the gas density.

For each cluster, we calculate $S_X$ projected along the $x$-, $y$-, and
$z$ axes according to Equation (\ref{e5sx}). We use the package {\tt
SPEX} to calculate the cooling function $\Lambda(T,Z)$ for the photon
energy band of $0.5$keV$<E<10$keV.  

In computing $S_X$, earlier SPH simulations had to exclude the
contribution from cold and dense gas particles because this may
overestimate the density of nearby hot X-ray emitting particles (Croft
et al. 2001; Kay et al.  2002). This can be a serious issue especially
for low resolution simulations in which the different gas phases are not
well resolved.  We believe that this is not a problem for our present
simulations because the gas densities are properly estimated on
individual cells, and also the mass and spatial resolutions are
significantly improved than those in their SPH simulations.

Then we directly fit ellipses to {$S_X(\vec{\theta})$}. We adopt the
fitting procedure of \cite{Jedrzejewski87} following
\cite{Kawahara10}. For each simulated cluster, we identify the ellipses
with semi-major axis $a_2$ fixed to $a_2/r_{200}=0.1$, 0.2, 0.3, and
0.4. Then the number of free parameters in the fitting procedure is
four; the axis ratio $q \equiv a_1/a_2$, direction of the major axis
$\Theta$ and the central position $\bm{X}_c$ {in the projected profile}.

\begin{figure}[tbp]
\begin{center}
\subfigure{
\FigureFile(60mm,60mm){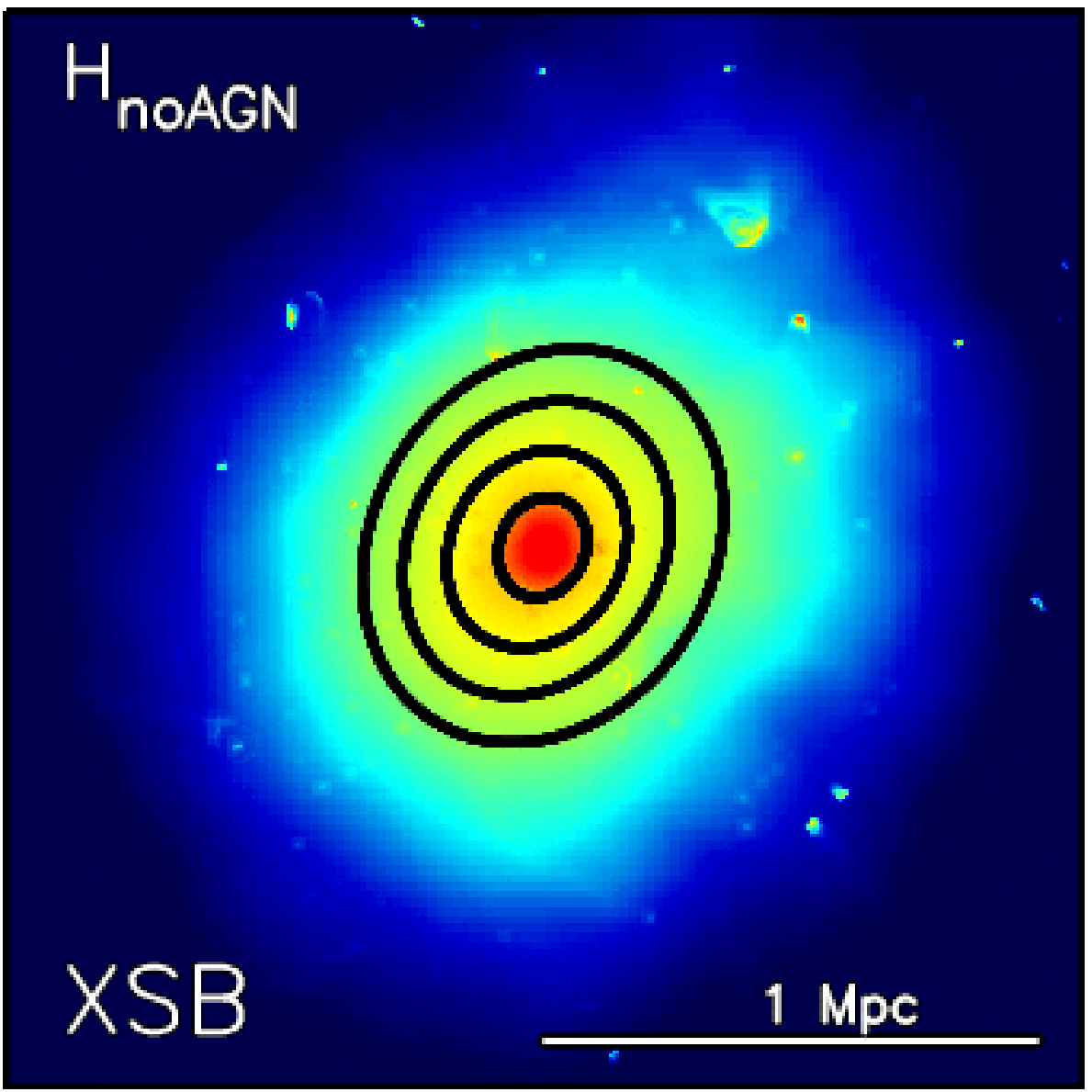}\qquad
\FigureFile(60mm,60mm){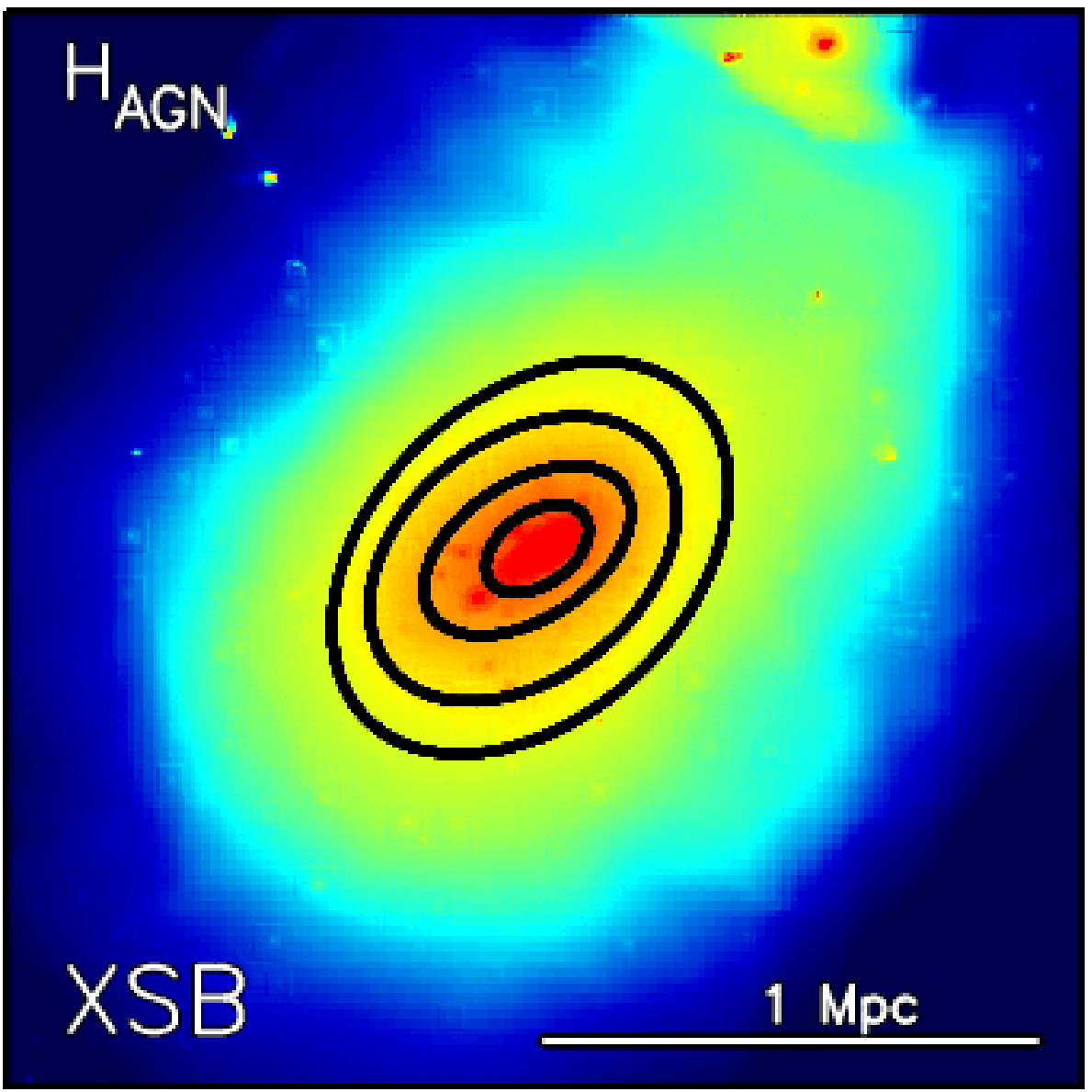}
}\\
\subfigure{
\FigureFile(60mm,60mm){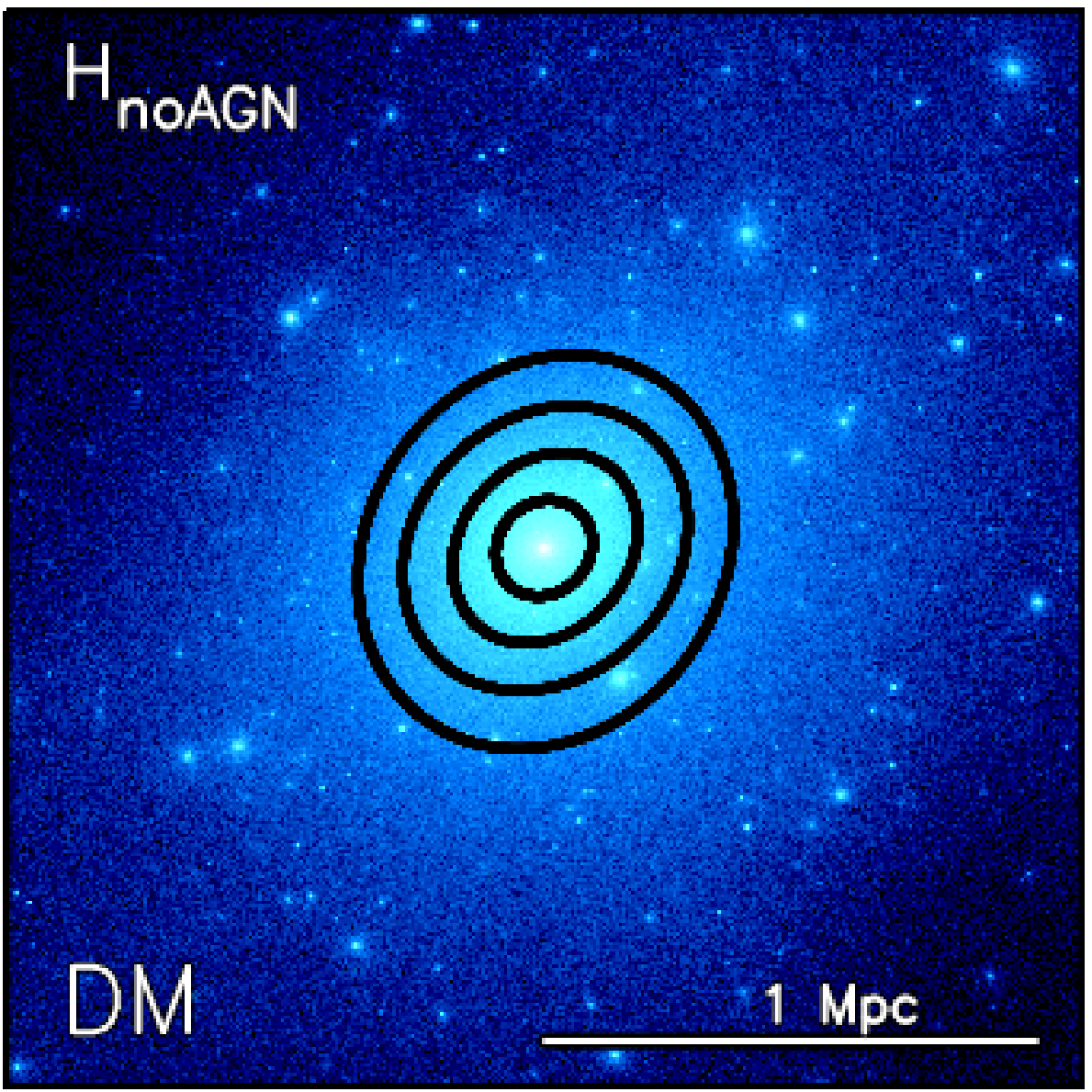}\qquad
\FigureFile(60mm,60mm){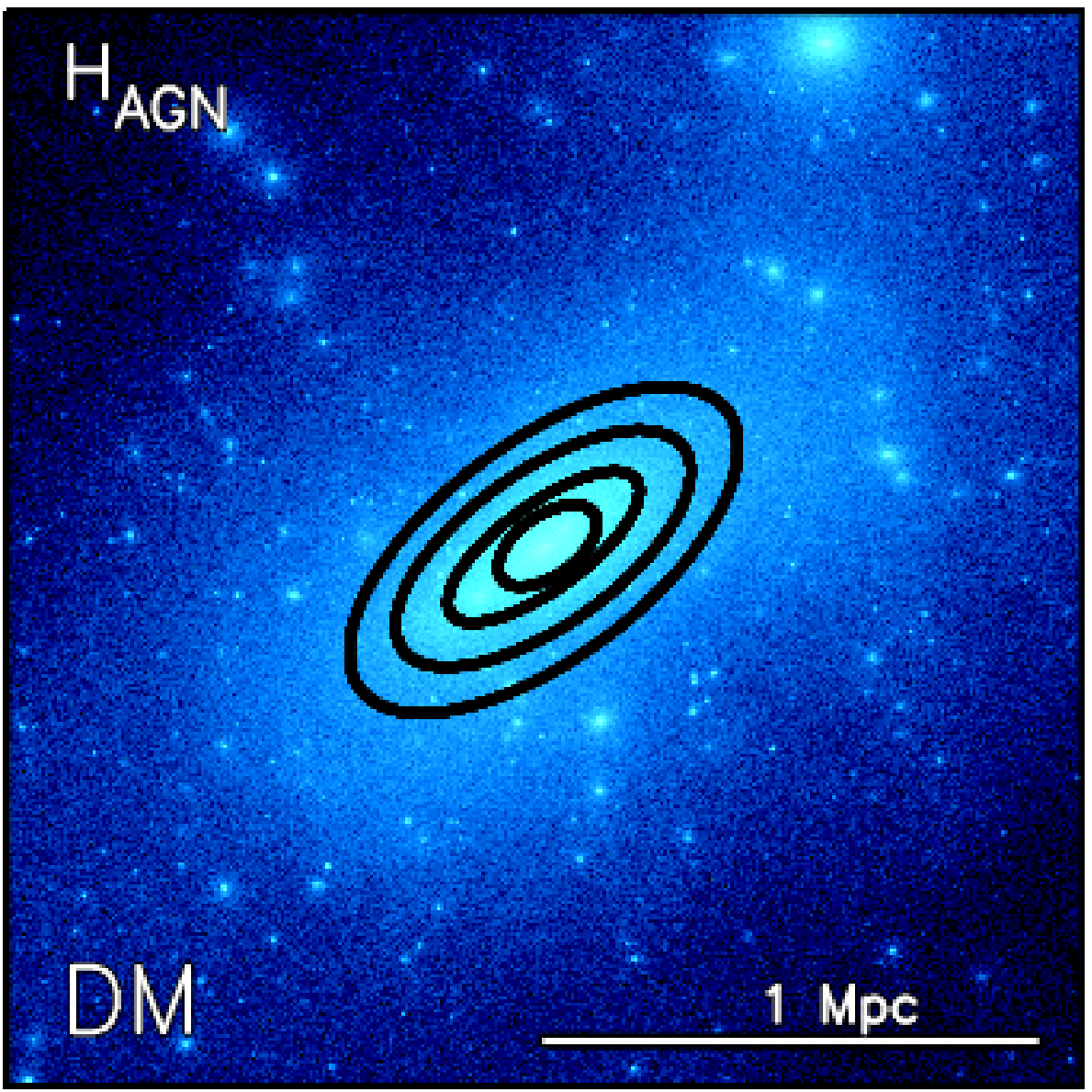}
}\\
\subfigure{
\FigureFile(60mm,60mm){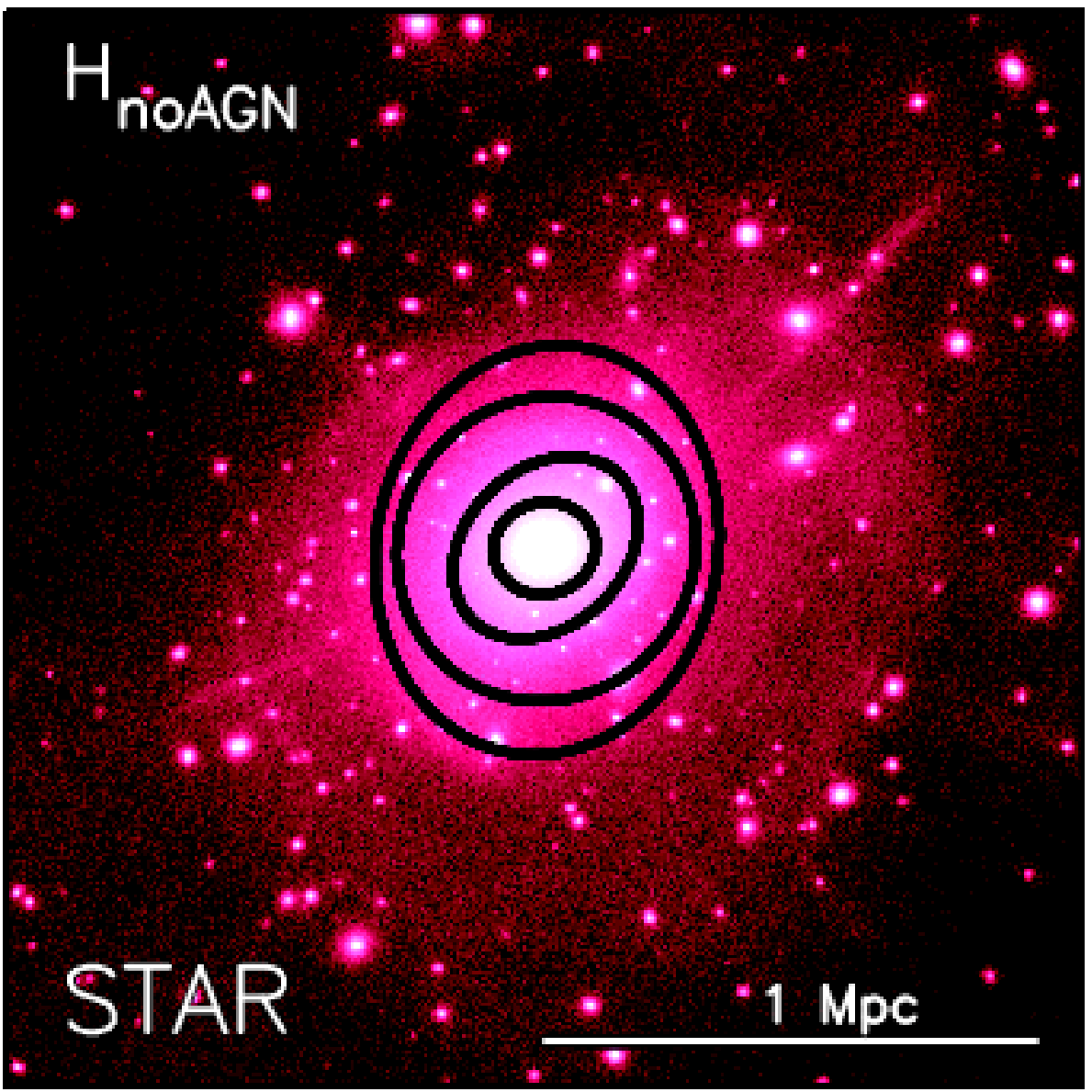}\qquad
\FigureFile(60mm,60mm){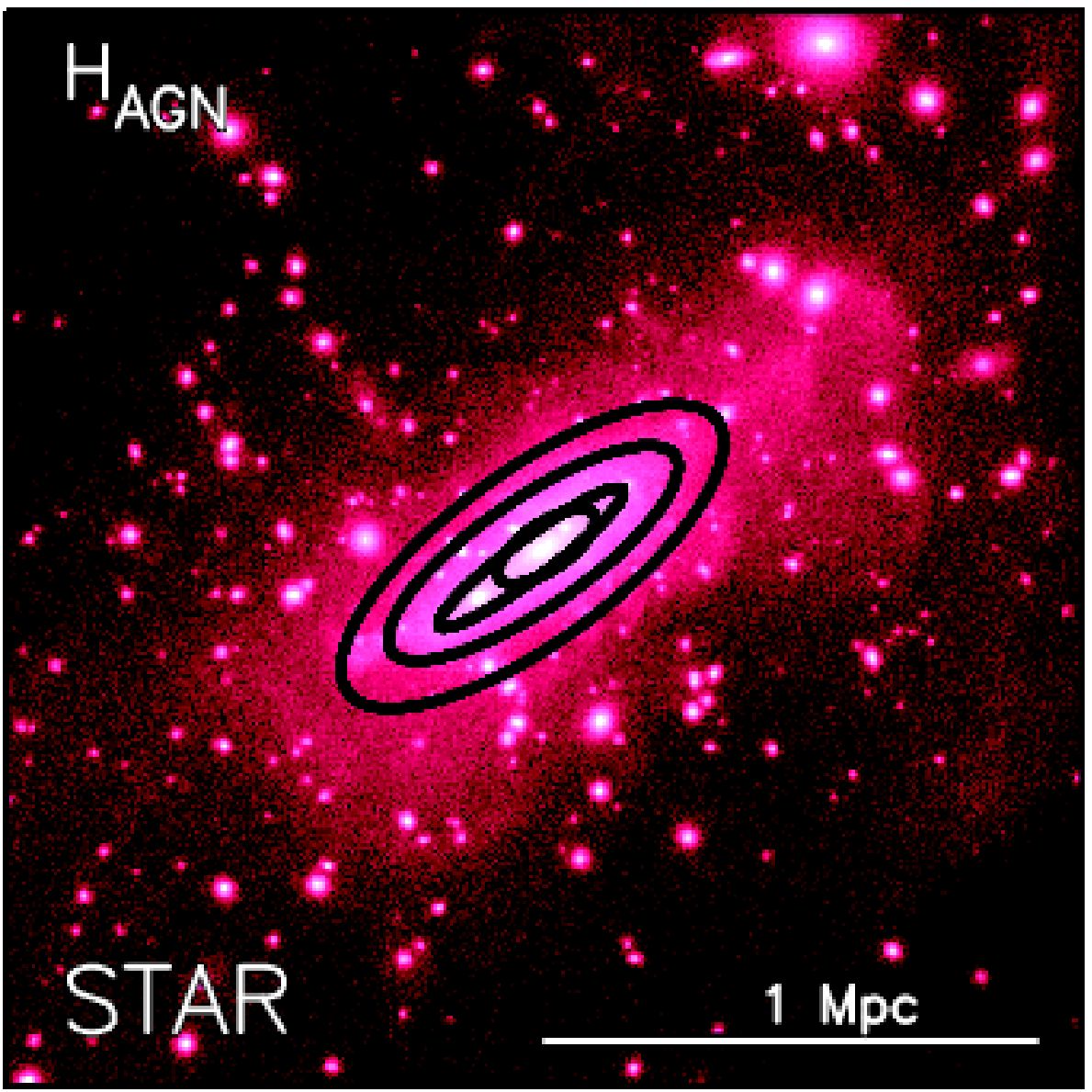}
}
\end{center}
\caption{X-ray surface brightness $S_X$ (top), dark matter density
distribution (middle), and stellar density distribution (bottom) of the
same cluster plotted in Figure \ref{f5dtwo} projected along the
$z$-axis; \HnoAGN\, (left) and \HAGN\, (right). For each panel, the
ellipses with the semi-major axis $a_2/r_{200}=0.1$, 0.2, 0.3{, and}
0.4 are plotted in black curves. The ellipses are obtained by direct
fitting for $S_X$, and by mass tensor for dark matter and stars.}
\label{f5map}
\end{figure}

Figure \ref{f5map} shows the distribution of $S_X$ (top), dark matter
density (middle) and stellar density (bottom), {projected along $z$-axis}
for the same cluster plotted in Figure \ref{f5dtwo}.  The left and right
panels correspond to \HnoAGN~ and \HAGN, respectively.  The black curves
illustrate the ellipses with the semi-major axis $a_2/r_{200}=0.1$, 0.2,
0.3, and 0.4.

{Comparing the ellipses for the three components in \HAGN, 
$S_X$ appears} more spherical, but the stellar density distribution
is more elongated, than the dark matter density distribution. The former
is consistent with the expectation that the gas distribution follows the
isopotential surfaces. We also note that the orientations of those
ellipses are similar among the three components, but they are not
necessarily concentric.

In contrast, the dark matter and stellar density distributions in
\HnoAGN\, are significantly more spherical than those in \HAGN.  This
indicates that the baryonic processes have strong impacts on the
non-sphericity of collisionless particles, even beyond the central
regions of galaxy clusters
\citep{Debattista08,Teyssier11,Bryan13,Butsky15,Cui16}.  In the next
section, we examine in more detail the statistics and correlation of the
projected axis ratios of the three components.

\section{Statistics and correlation of the non-sphericity of gas, star{,}
 and dark matter distribution \label{sec:statistics}}

\subsection{Effect of baryonic physics on  dark
 matter distribution \label{subsec:baryon-DM}}

As shown in the previous section, the baryon physics affects the
non-sphericity of dark matter distribution {significantly, perhaps} more than what we naively
expect. We first examine this surprising result more quantitatively
below.

\begin{figure}[tbh]
\begin{center}
\subfigure{
\FigureFile(70mm,70mm){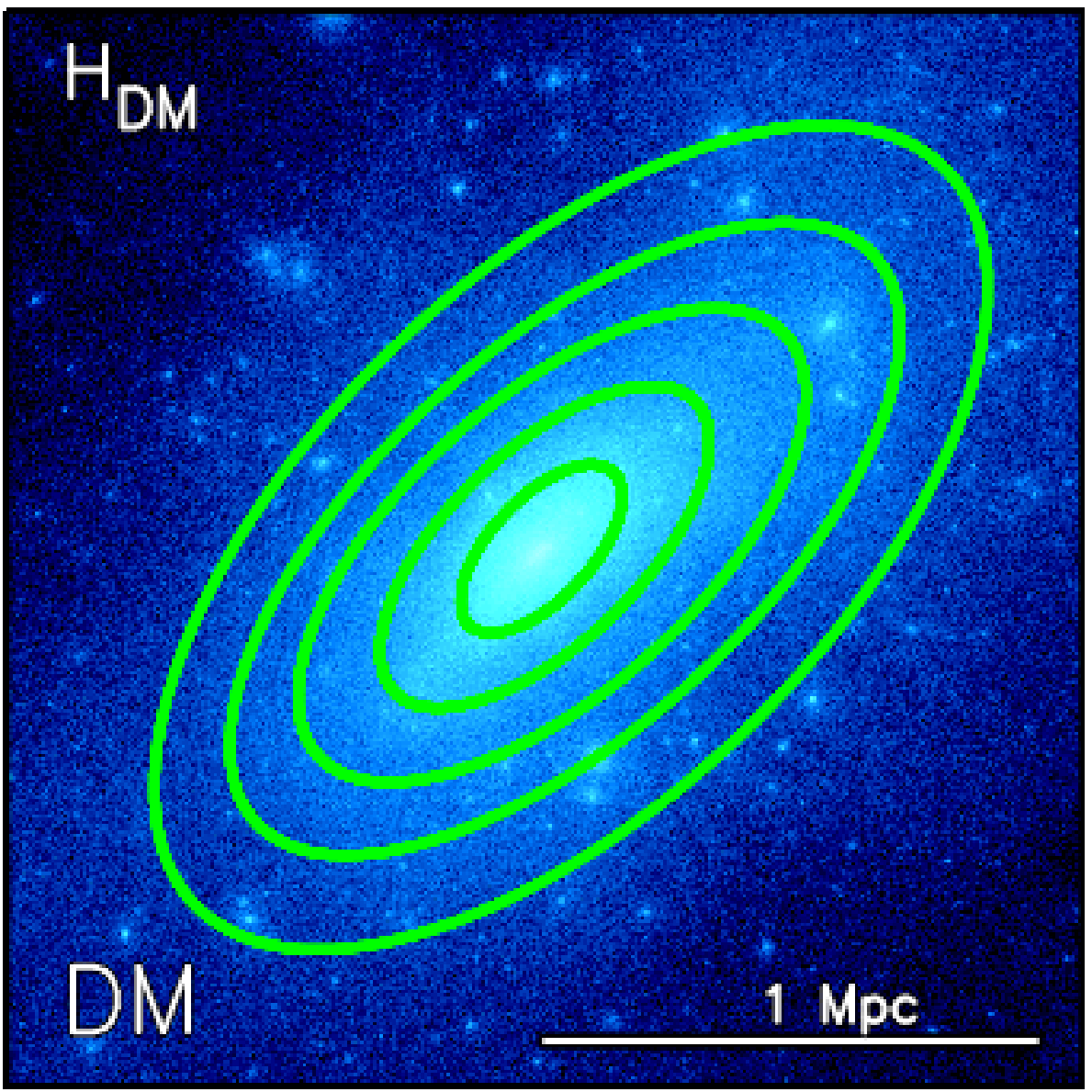}
\qquad
\FigureFile(70mm,70mm){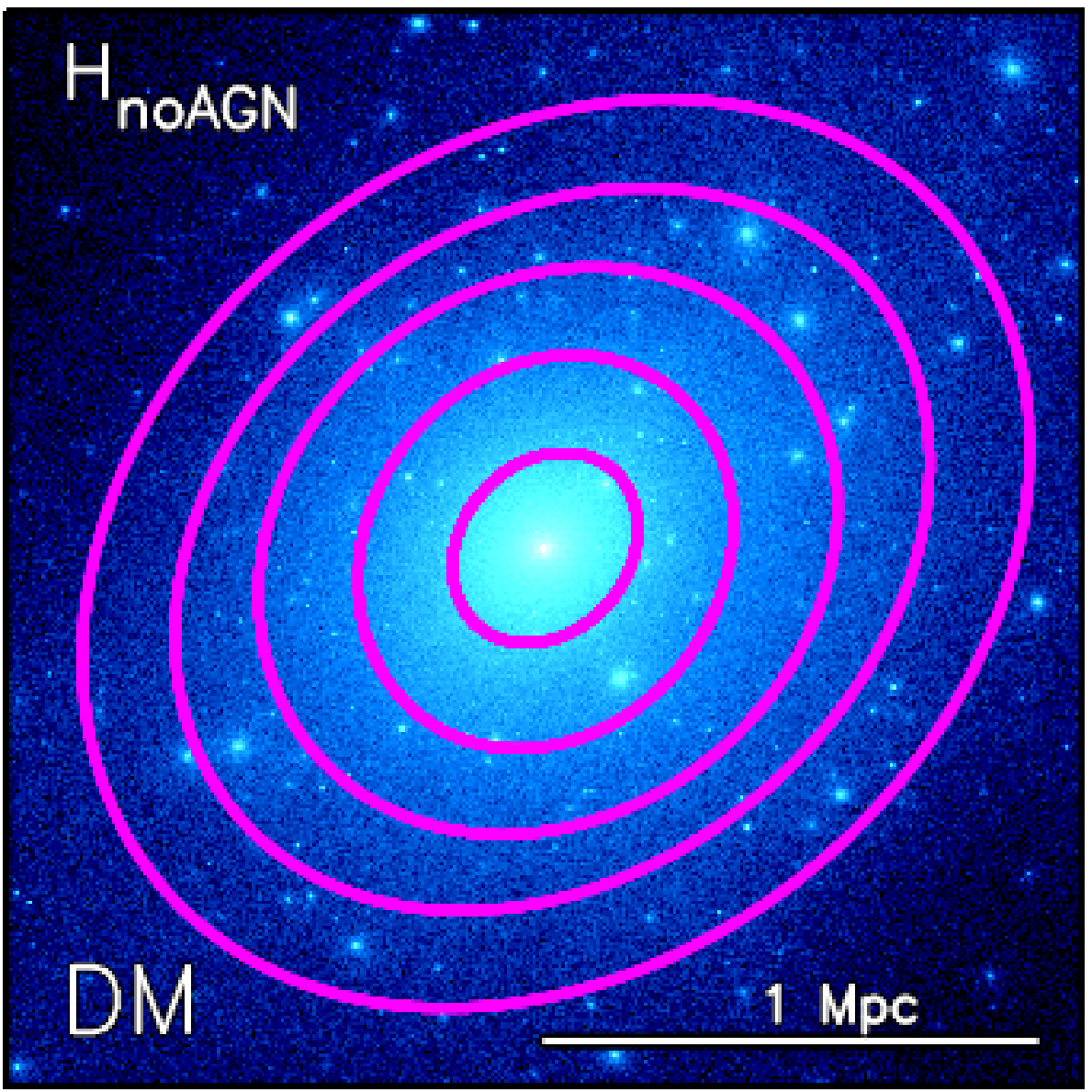}
}\\
\subfigure{
\FigureFile(70mm,70mm){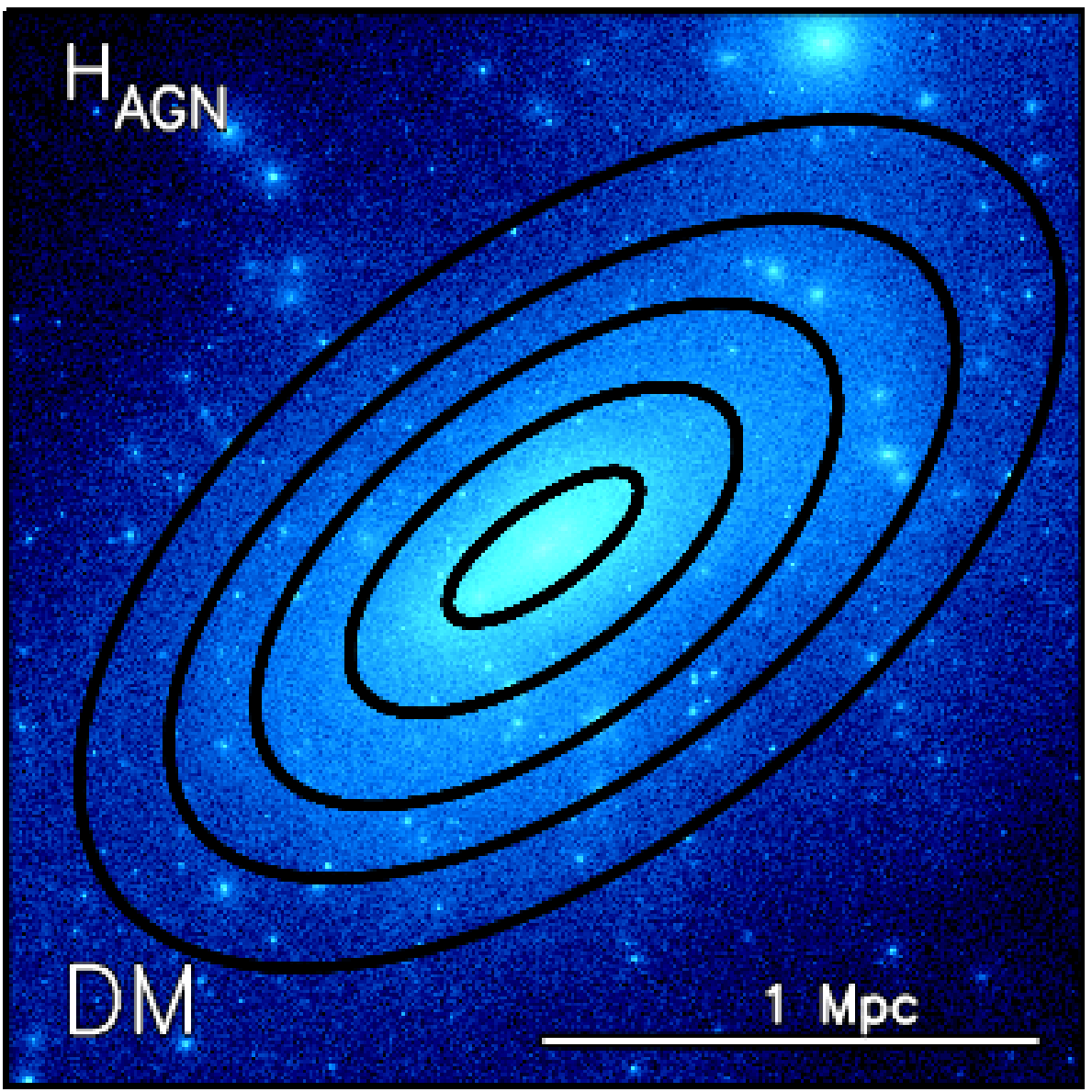}
\qquad
\FigureFile(70mm,70mm){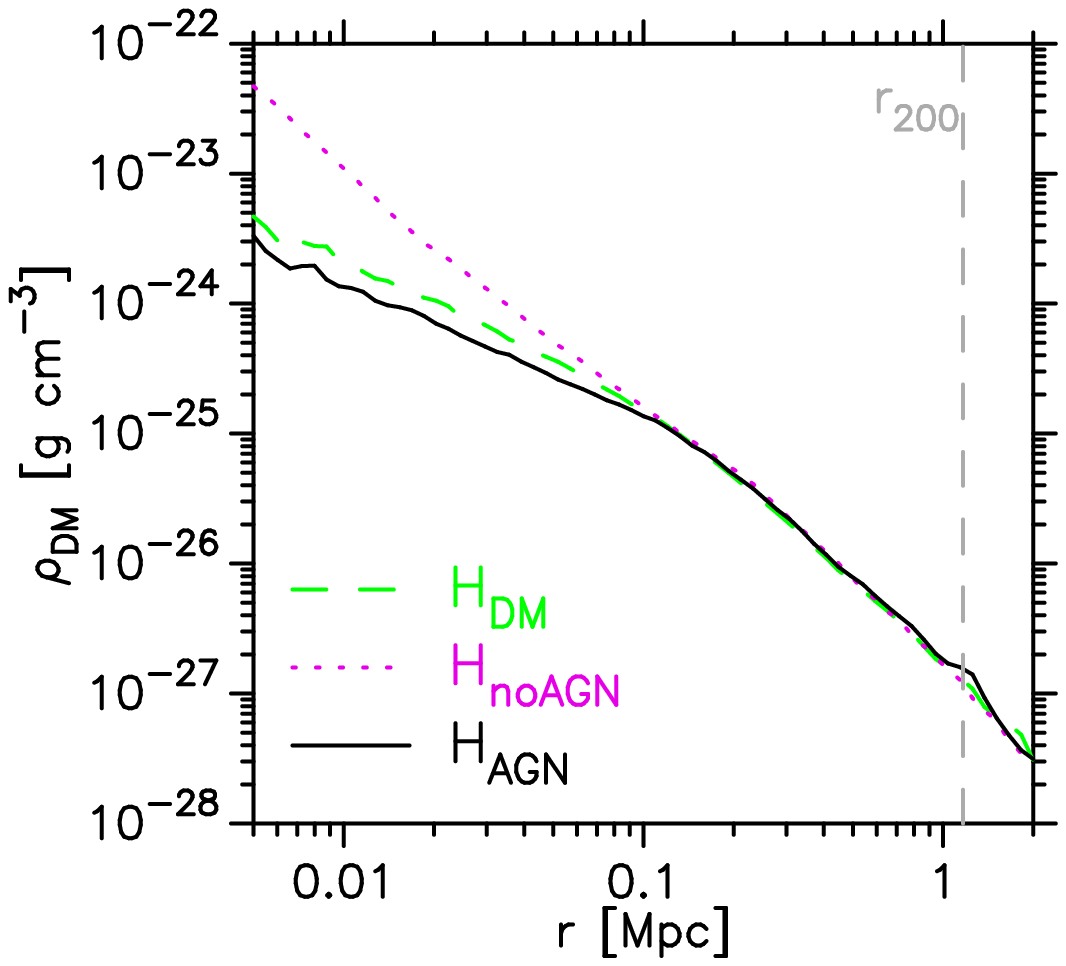}
}
\end{center}
\caption{Dark matter density distribution of the same cluster in Figure
\ref{f5dtwo} projected along the $z$-axis of the simulation for three
simulations; \HDM~ (upper-left), \HnoAGN~ (upper-right) and \HAGN~
(lower-left). The fitted ellipses with the semi-major axis
$a_2/r_{200}=0.2$, 0.4, 0.6, 0.8, 1.0 are plotted in black curves. The
corresponding dark matter density profiles are shown in the lower-right
panel; \HDM~(green), \HnoAGN~(magenta) and \HAGN~(black). The dashed
vertical line indicates $r_{200}$ in \HAGN. For \HDM, the density
profile is multiplied by a factor of
$1-\Omega_{b,0}/\Omega_{m,0}\approx0.83$.}  \label{f5dmdist}
\end{figure}

Figure \ref{f5dmdist} plots the projected dark matter distribution 
{of}
the same cluster in Figures \ref{f5dtwo} and \ref{f5dtw}
{, extracted from} \HDM\,
(upper-left), \HnoAGN\,(upper-right), and \HAGN\,(lower-left).  The
ellipses {computed} from the mass tensor corresponding to the
semi-major axis of $a_2/r_{200}=0.2$, 0.4, 0.6, 0.8, and 1.0 are plotted
in solid curves.  The lower-right panel shows the {\it spherically
averaged} profiles of dark matter density for \HDM\,(green),
\HnoAGN\,(magenta) and \HAGN\,(black). Because the three {runs} adopt
the same value for the matter density $\Omega_{m,0}$ {and the run
\HDM~ does not consider the baryon component, 
the dark matter mean density is different for this run;} $\Omega_{\rm
CDM,0}=\Omega_{m,0}$ for \HDM, but $\Omega_{\rm
CDM,0}=\Omega_{m,0}-\Omega_{b,0}$ for \HAGN~ and \HnoAGN. Thus 
the density profile of \HDM~ in the lower-right panel is
multiplied by a factor of $1-\Omega_{b,0}/\Omega_{m,0}(\approx0.83)$.

As is shown in the lower-right panel, the spherically averaged dark
matter density profiles for $r>0.1$Mpc are almost the same for the three
{runs}.
{On the contrary,} the inner profiles are
significantly affected by baryon physics.  The effect is particularly
strong for \HnoAGN~ that neglects the AGN feedback; the gas over-cooling
pulls gas and stars towards the center, and then dark matter particles
in the outer region fall into the central region.  The AGN feedback,
however, suppresses the gas over-cooling and flattens the inner density
profile
\citep{Peirani2008,Dubois10,Dubois11,Teyssier11,Martizzi2013,Peirani2016}.
Thus the dark matter density profile of \HAGN~ turns out to be fairly
close to the result for dark matter only simulation (\HDM).
Qualitatively speaking, the above tendency may be consistent with what
we expect.

\begin{figure}[tbp]
\begin{center}
\FigureFile(70mm,70mm){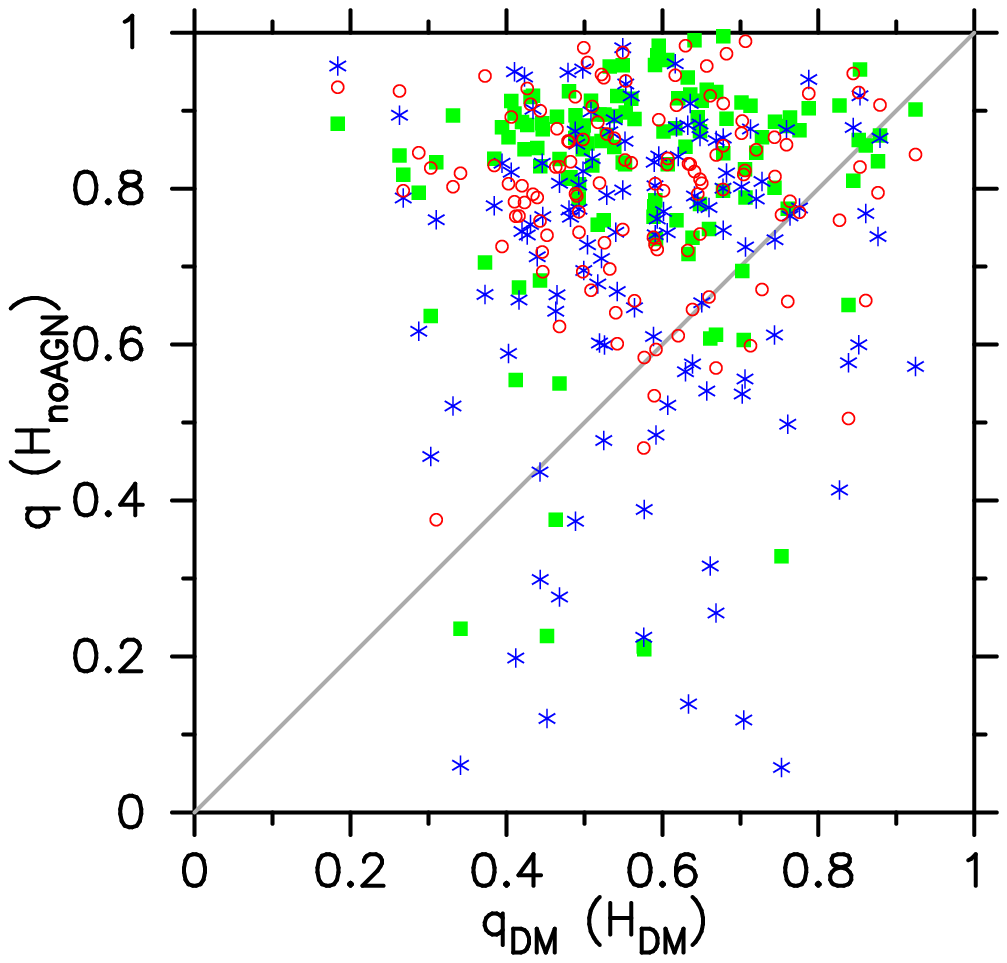}
\quad
\FigureFile(70mm,70mm){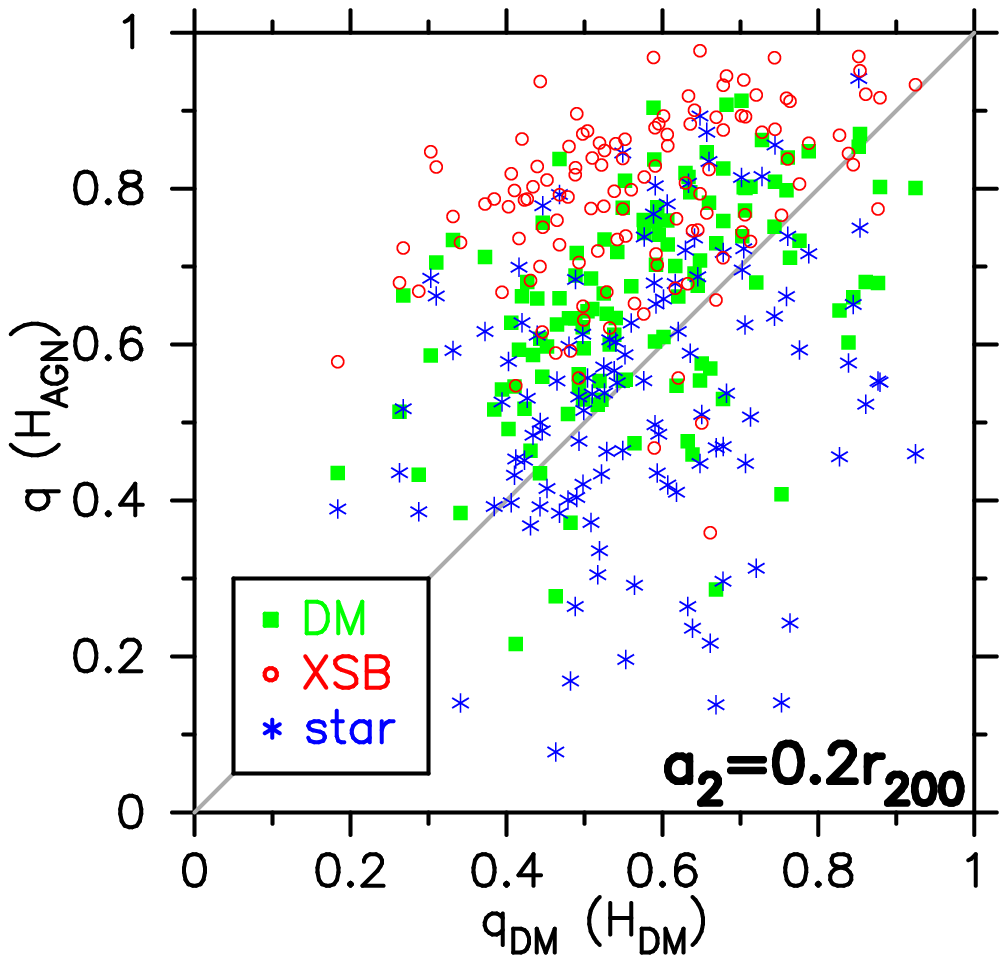}
\quad
\FigureFile(70mm,70mm){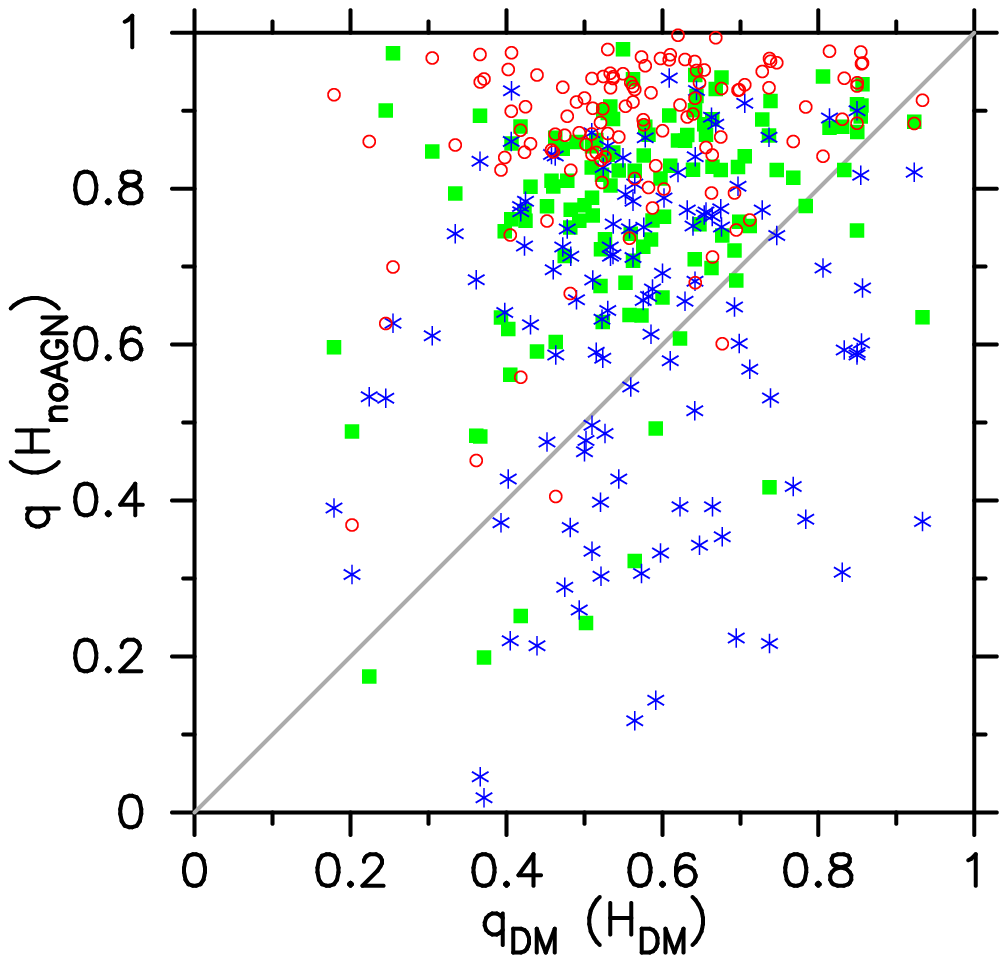}
\quad
\FigureFile(70mm,70mm){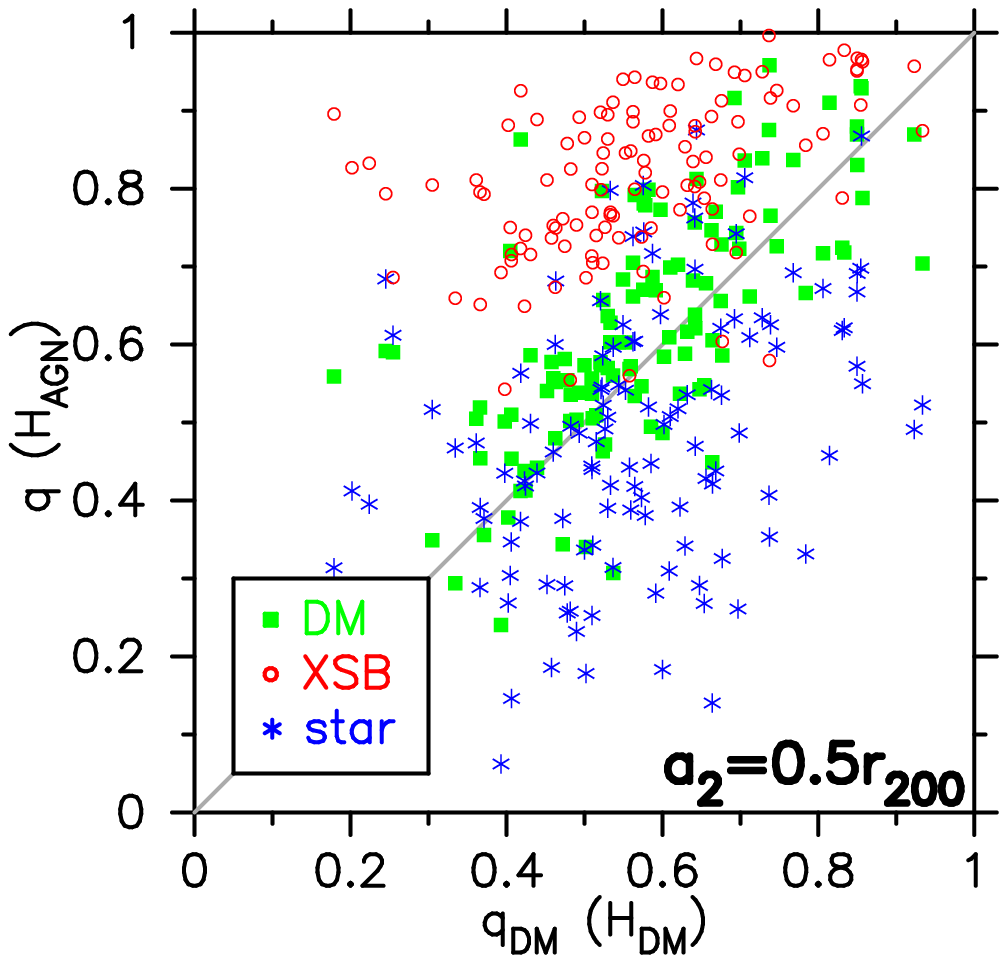}
\quad
\FigureFile(70mm,70mm){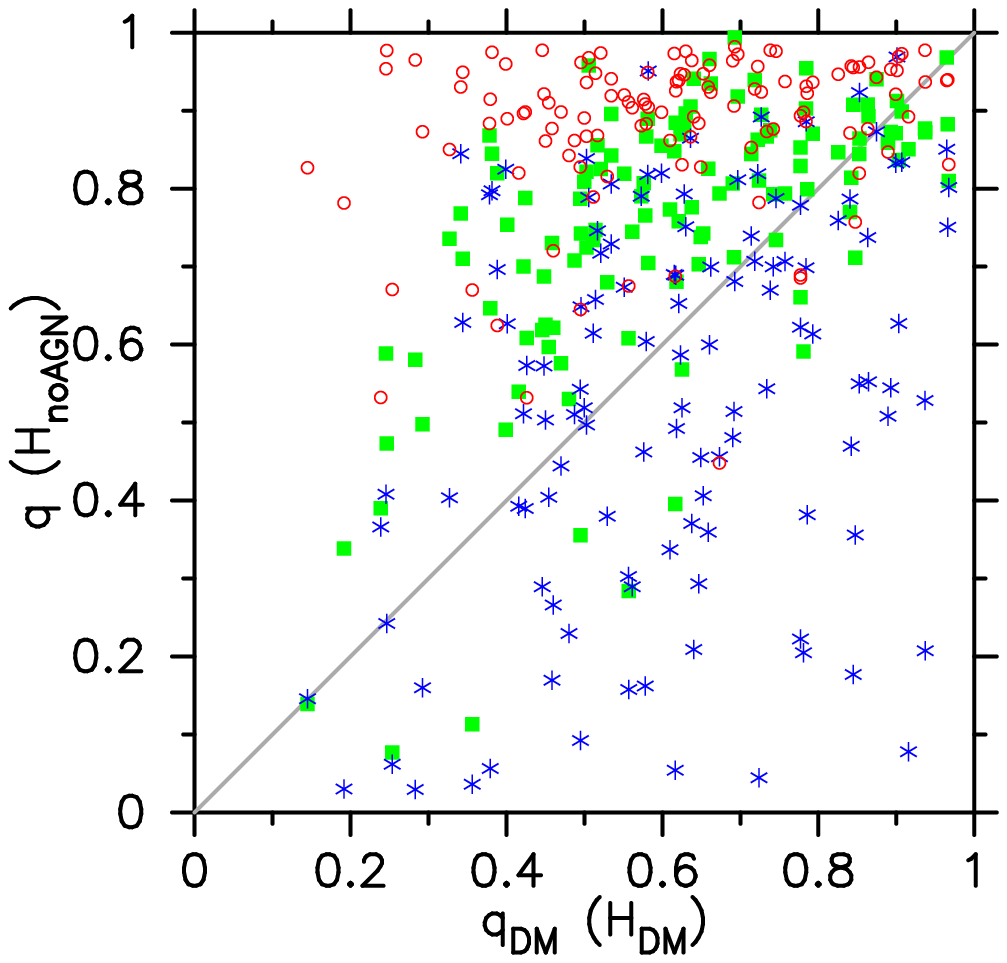}
\quad
\FigureFile(70mm,70mm){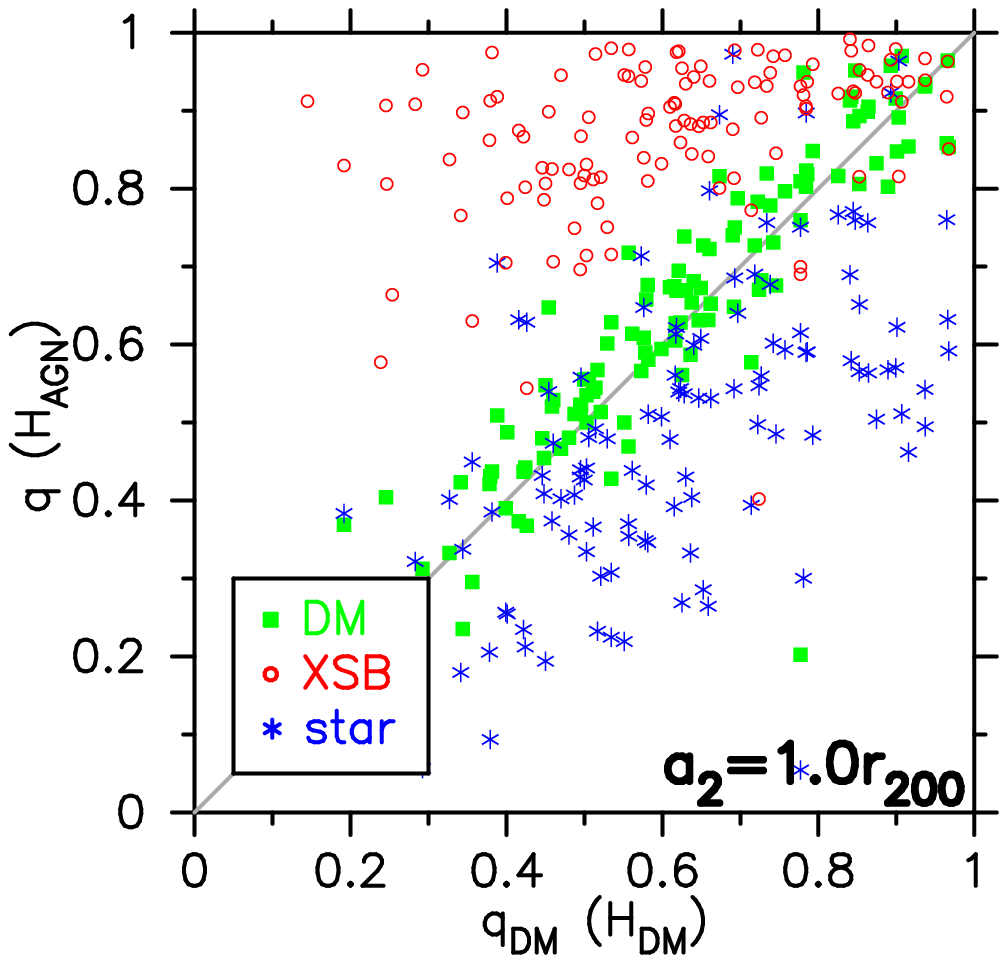}
\end{center}
\caption{Projected axis ratios of simulated clusters in \HAGN\, ({\it
left}) and \HnoAGN\, ({\it right}) against their counterparts in \HDM;
$a_2=0.2r_{200}$ ({\it upper}), $a_2=0.4r_{200}$ ({\it middle}), and
$a_2=r_{200}$ ({\it lower}). } \label{fig:q-comparison}
\end{figure}

{In order to make more quantitative comparison, we show in Figure
\ref{fig:q-comparison}} the projected axis ratios $q$ for the 40
simulated clusters (viewed from three different line-of-sights) against
their counterparts in \HDM~ at $a_2=0.2r_{200}$ ({\it top}),
$a_2=0.5r_{200}$ ({\it middle}), and $a_2=r_{200}$ ({\it bottom}).  We
note that $0.2r_{200}$ and $0.5r_{200}$ roughly correspond to the mass
scales of $M_{2500}$ and $M_{500}$.

Although the lower-right panel in Figure \ref{f5dmdist} may be
interpreted that the baryonic effect is largely {absent} for $r>0.1$Mpc,
Figure \ref{fig:q-comparison} indicates that it is not the case {as long
as} the non-sphericity {is concerned}.  The dark matter distribution in
\HnoAGN~ is significantly rounder than, and not so correlated with, that
in \HDM. This result would be caused by the unrealistic gas over-cooling
in the absence of AGN feedback.  With the AGN feedback, $q_{\rm
DM}$(\HAGN) is roughly correlated with $q_{\rm DM}$(\HDM) but with large
scatters.  The deviation between $q_{\rm DM}$(\HAGN) and $q_{\rm
DM}$(\HDM) becomes weaker for outer regions, but still {clearly} exists
even around the virial radius, $a_2=r_{200}$.  This implies that the
baryon processes around the inner region of galaxy clusters affect the
overall shape of their dark matter halos, {which is indeed surprising.}

\begin{figure}[tbp]
\begin{center}
\FigureFile(50mm,50mm){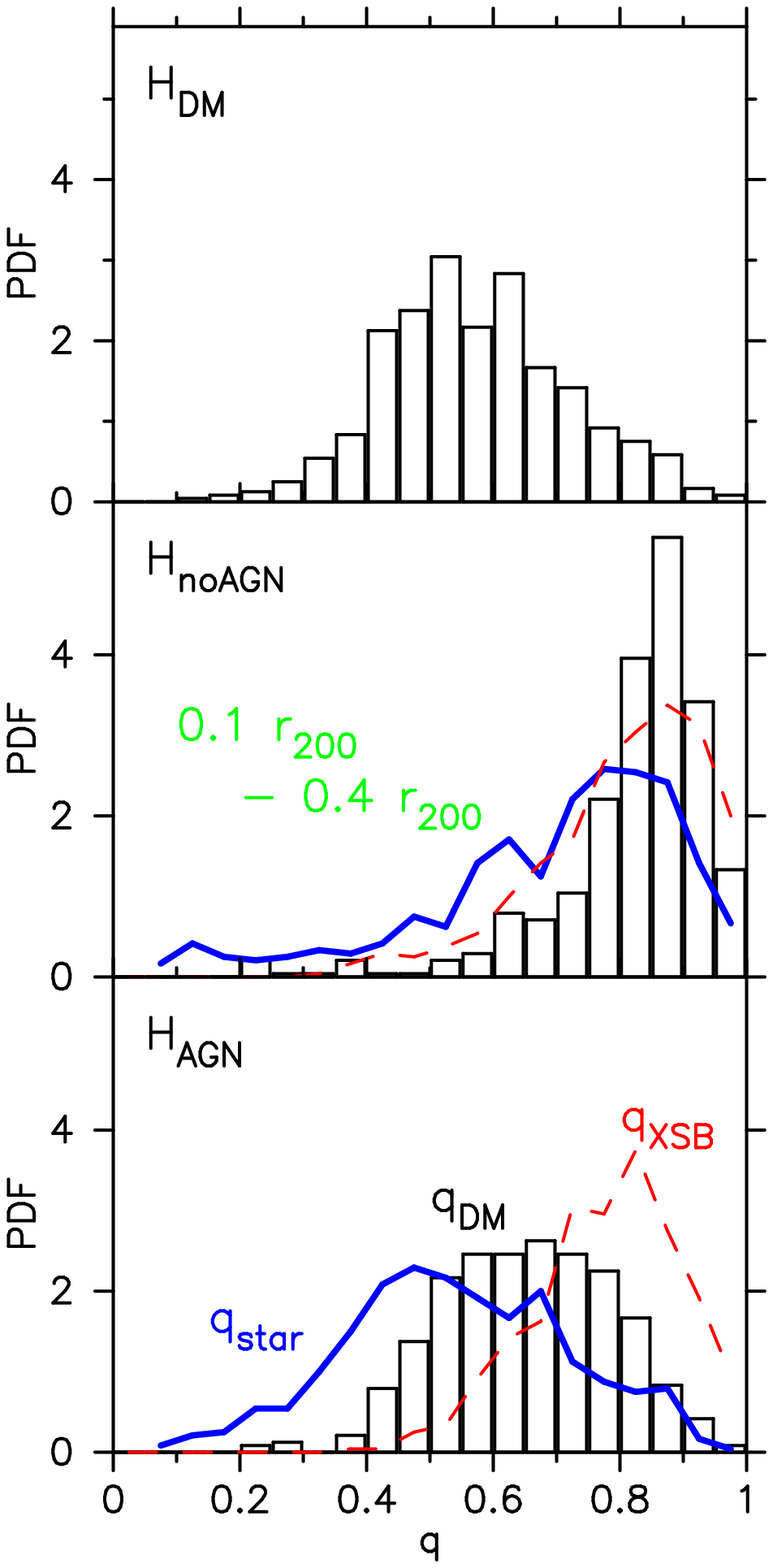}
\quad
\FigureFile(45mm,45mm){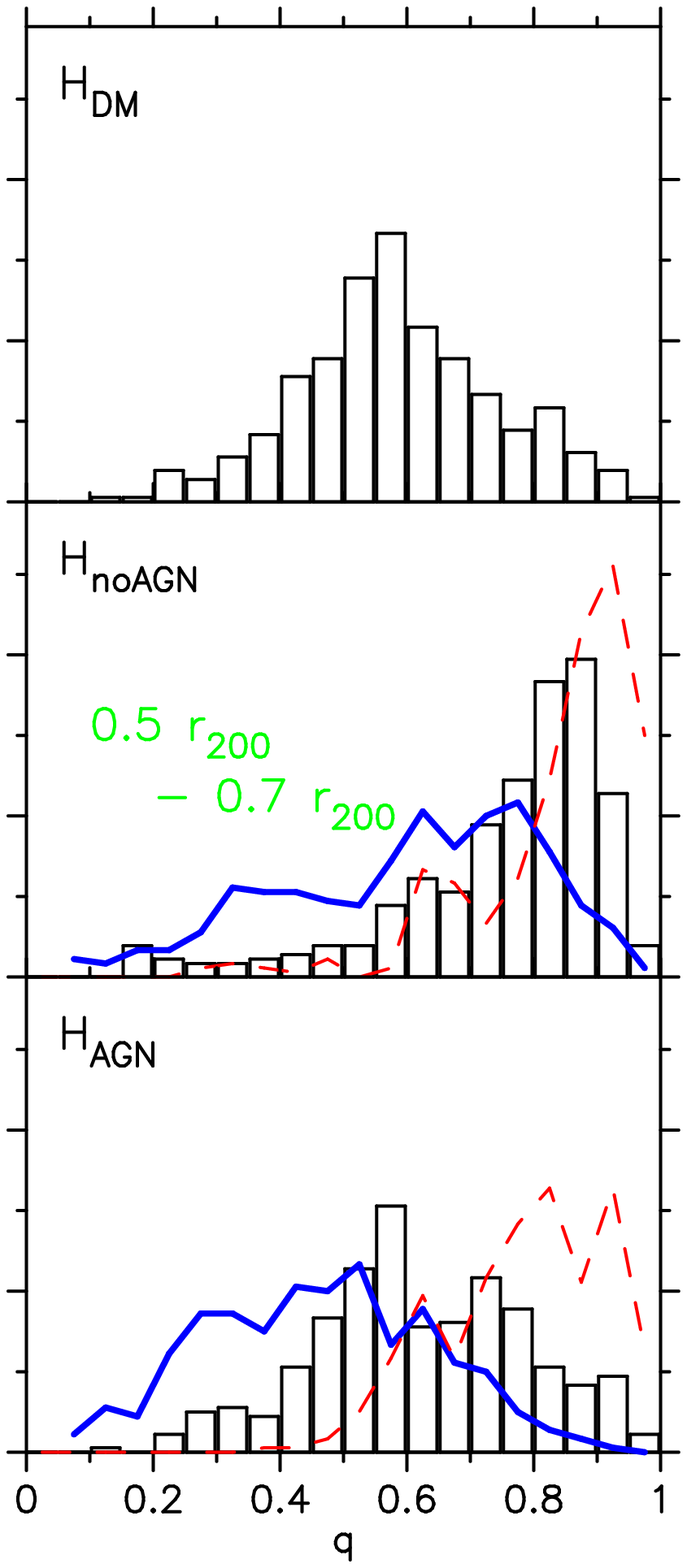}
\quad
\FigureFile(45mm,45mm){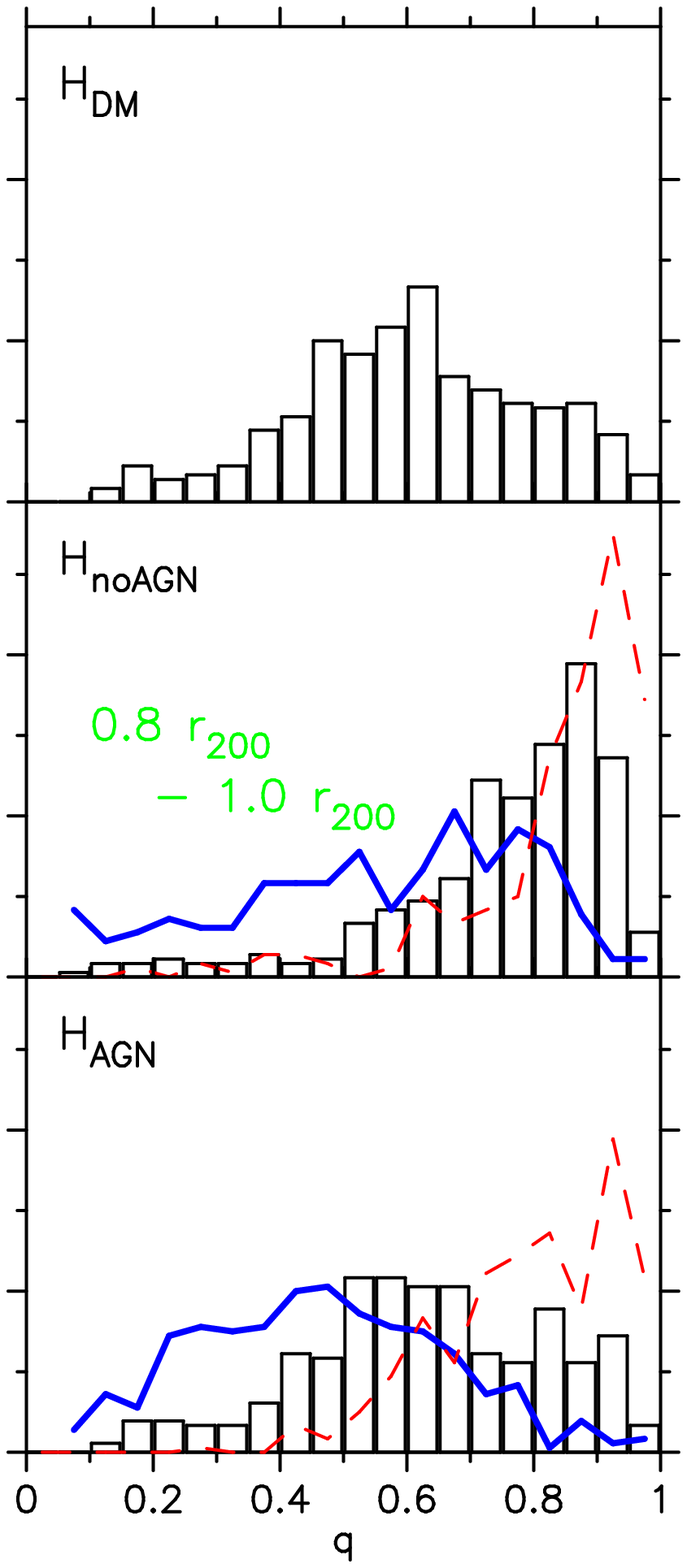}
\end{center}
\caption{PDFs of projected axis ratio of dark matter density
distribution for the 40 simulated cluster along three lines-of-sight
($x$-, $y$-, and $z$-axes of the simulation). For each cluster, ellipses
with the semi-major axis $a_2/r_{200}=0.1$, 0.2,..., 0.9, and 1.0 are
fitted by the mass tensor, and the PDF is calculated for the ellipses
with $a_2=0.1$ - 0.4 $r_{200}$ (left), 0.5 - 0.7 $r_{200}$ (middle) and
0.8 - 1.0 $r_{200}$ (right). The same analysis is performed for the
three kinds of simulations; \HDM\, (top), \HnoAGN\, (middle) and \HAGN\,
(bottom). For \HAGN\, and \HnoAGN\,, the PDFs of the axis ratios of
stellar density and XSB distributions 
are also plotted in blue solid and red dashed lines, respectively.
\label{f5qdm}}
\end{figure}

To see statistically the effect of baryons on the non-sphericity of dark
matter halos, we compute the probability density function (PDF) of the
projected axis ratios.  Figure \ref{f5qdm} shows the result for \HDM\,
(top), \HnoAGN\, (middle), and \HAGN\, (bottom).  To compensate for the
small number of clusters, we plot the combined PDFs at
$a_2/r_{200}=0.1$, 0.2, 0.3, and 0.4 (left), 0.5, 0.6, and 0.7 (middle)
and 0.8, 0.9, and 1.0 (right); the number of samples is 480, 360, and
360, respectively.

{At the scale of $0.1 \le a_2/r_{200} \le 0.4$}, the PDF of $q_{\rm
DM}$ for \HDM\. (top)
is basically consistent with the result of \citet{Suto16b} for
$M_{2500}$ ($r_{2500}\approx0.2 r_{200}$), although the number of dark
matter halos is much smaller {due to the smaller box size of the
Horizon simulation}. The PDF of $q_{\rm
DM}$ for \HnoAGN\, (middle panel of Figure
\ref{f5qdm}) is systematically shifted towards the right, relative to
that of \HDM; the dark matter density distribution is significantly
rounder.  If the AGN feedback is included (bottom), the density
distribution is only slightly rounder than that of \HDM.

For reference, we plot the PDFs {of $q_{\rm star}$ (blue solid
lines) and $q_{\rm XSB}$ (red dashed lines) for \HnoAGN\, and \HAGN\, in
Figure \ref{f5qdm}.} For both runs, the PDF of $q_{\rm star}$
is slightly shifted to the left (less spherical) compared to that of
dark matter in each {run}. This is partly because the mass tensor,
equation (\ref{eq:masstensor}), is sensitive to substructures, i.e.,
galaxies at large radii. The PDFs {of $q_{\rm star}$, $q_{\rm XSB}$,
and $q_{\rm DM}$ for \HnoAGN\, are all systematically shifted to the
right, relative to those for \HAGN\, due to the over-cooling.}

The PDFs in Figure \ref{f5qdm} statistically confirm the visual
impression of Figures \ref{f5map} and \ref{f5dmdist}, and indicate
that the baryon processes around the central region have a strong impact
on the shape of galaxy clusters represented by dark matter even at their
virial radii.  This is surprising given that the spherically-averaged
density {profiles are} roughly the same {among} the three {runs}, and
that the dark matter occupies a much larger mass fraction ($\approx$ 80
\%) than gas and stars.

\cite{Kazantzidis04} showed that dark matter halos in a simulation
without radiative gas cooling are much less spherical than those in a
simulation with cooling (both simulations do not include AGN feedback)
up to the virial radius ($\approx r_{200}$). {Their} result also
indicates that the shape of dark matter halos is strongly influenced by
the {detail} of baryonic processes implemented in simulations. (Note that,
since they adopted a definition of mass tensor {somewhat different} from
equation (\ref{eq:masstensor}), their results should not be
quantitatively compared with ours.)  {We admit that the} {reason why}
{the baryon physics} {influences} {the shape of the outer region of dark matter halo
is not yet fully} {explained, and would like to defer the problem to the future study.}

The above results indicate that {dark matter alone simulations cannot be used
even to predict the non-sphericity of dark matter distribution.}

\subsection{Mass- and Radial-Dependence of the axis ratio}

\begin{figure}[tbp]
\begin{center}
\FigureFile(80mm,80mm){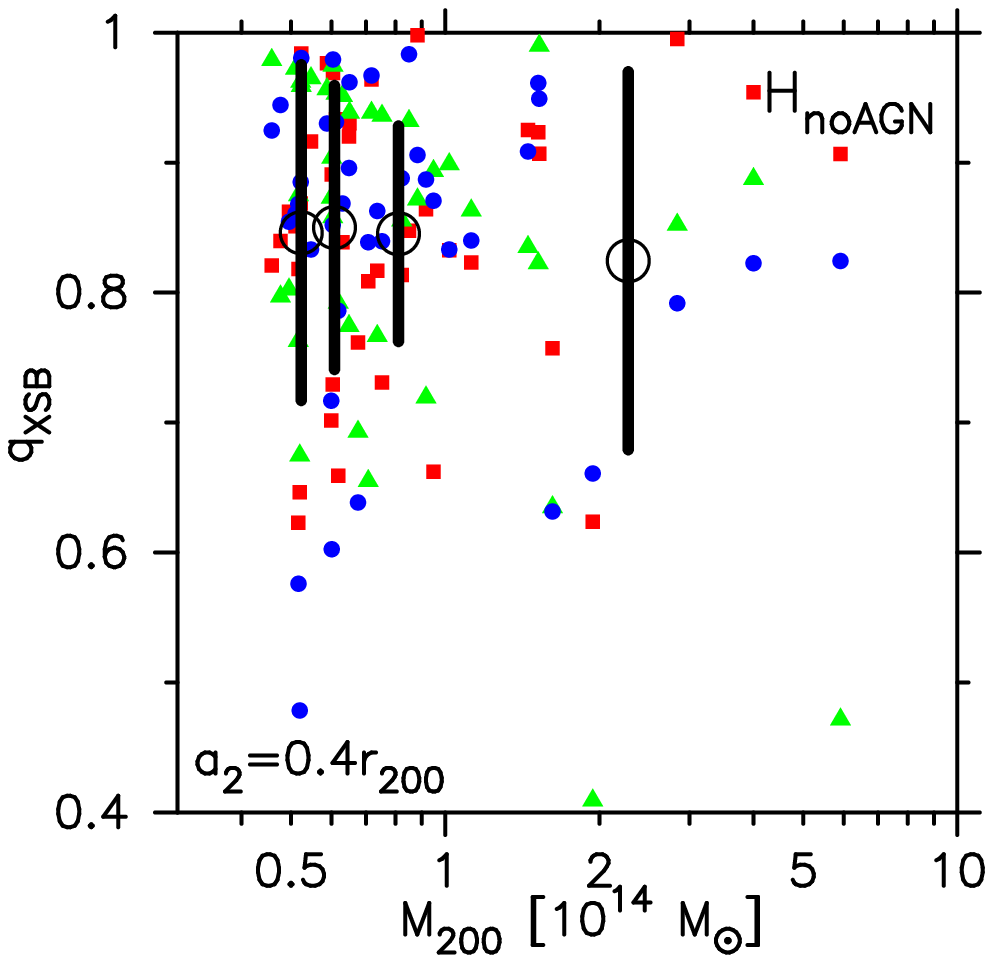}
\qquad
\FigureFile(80mm,80mm){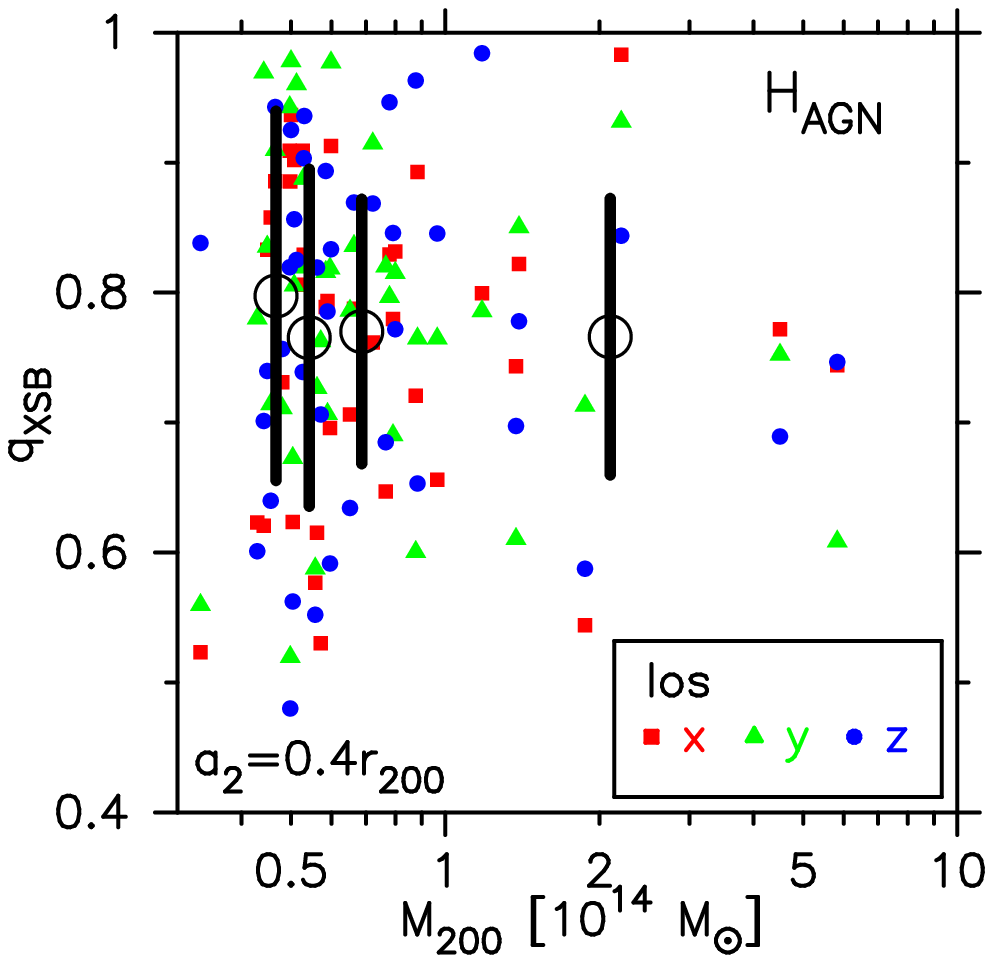}
\end{center}
\caption{Projected axis ratio $q$ of the X-ray surface brightness with
$a_2=0.4r_{200}$, against $M_{200}$ of each simulated cluster, for
\HnoAGN~ (left) and \HAGN~ (right). Each symbol indicates the result for a
simulated cluster. The axis ratio $q$ of each cluster is measured along
the three different lines-of-sight, and indicated in different symbols;
$x$-axis (red square), $y$-axis (green triangle), $z$-axis (blue
circle). The black circles show the averaged values of $q$ over every 10
of 40 clusters, and the black lines indicate the corresponding standard
deviation.}  \label{f5mq}
\end{figure}

Next we consider {the mass dependence of} the axis ratio $q$ of {SXB} of simulated galaxy
clusters.  Figure \ref{f5mq} plots $q_{\rm
XSB}$ at $a_2=0.4r_{200}$ for each simulated cluster along three
lines-of-sight against its $M_{200}$ for \HAGN\, (left) and \HnoAGN\,
(right).  For each cluster, the measured $q_{\rm XSB}$ along the $x$-,
$y$-, and $z$-axis are plotted in red squares, green triangles, and blue
circles, respectively. Even for the same cluster, $q$ can be quite
different depending on the lines-of-sight; for example, if a cluster is
elongated along the $z$-axis, $q$ {for the projection} along $x$- or
$y$-axis is much smaller than that along the $z$-axis.

We divide the 40 clusters into four groups of 10 clusters in the
decreasing order of $M_{200}$. The mean value of $q_{\rm XSB}$ and the
standard deviation are indicated by the black circle with error-bar in
Figure \ref{f5mq}. The mean $q_{\rm XSB}$ in \HnoAGN\, ($\sim$0.84) is
slightly larger then that in \HAGN\, $\sim$0.78. This reflects the
rounder shapes of the dark matter halos in \HnoAGN, but the difference
of $q_{\rm XSB}$ between the two simulation runs is smaller than that of
dark matter.

\begin{figure}[tbp]
\begin{center}
\FigureFile(80mm,80mm){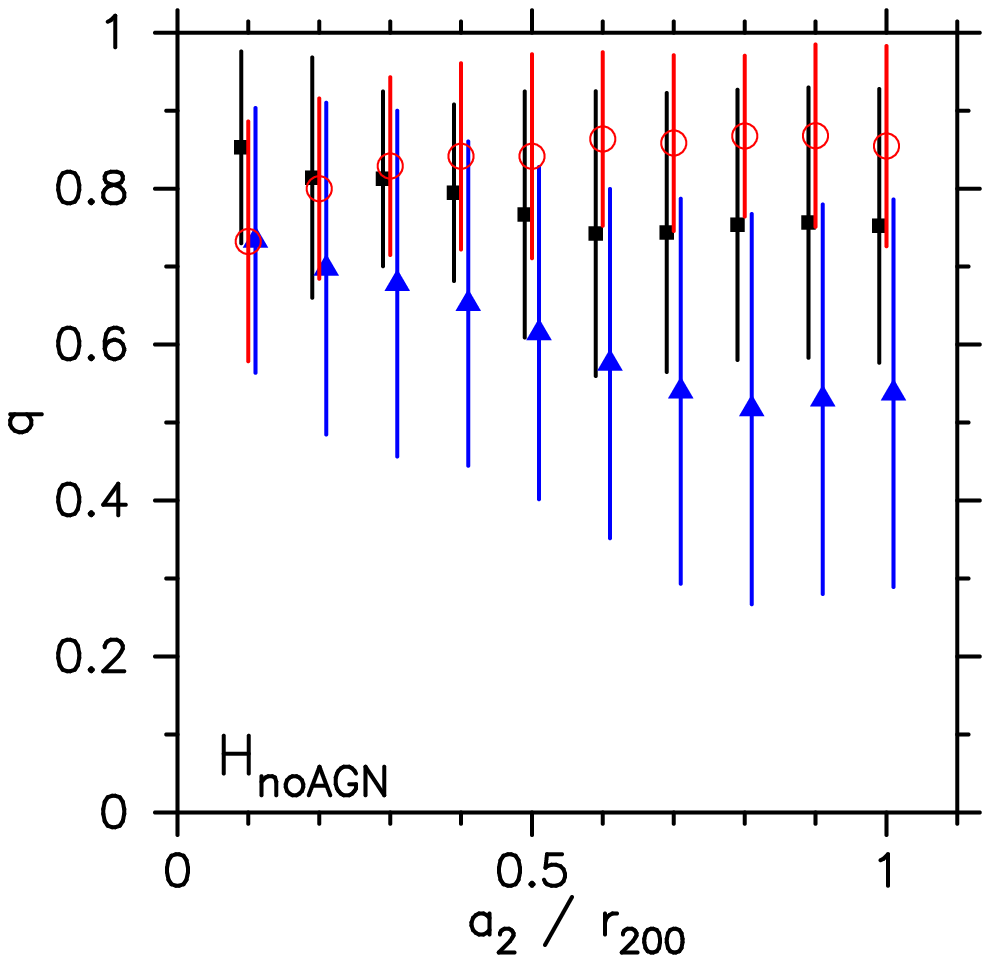}
\qquad
\FigureFile(80mm,80mm){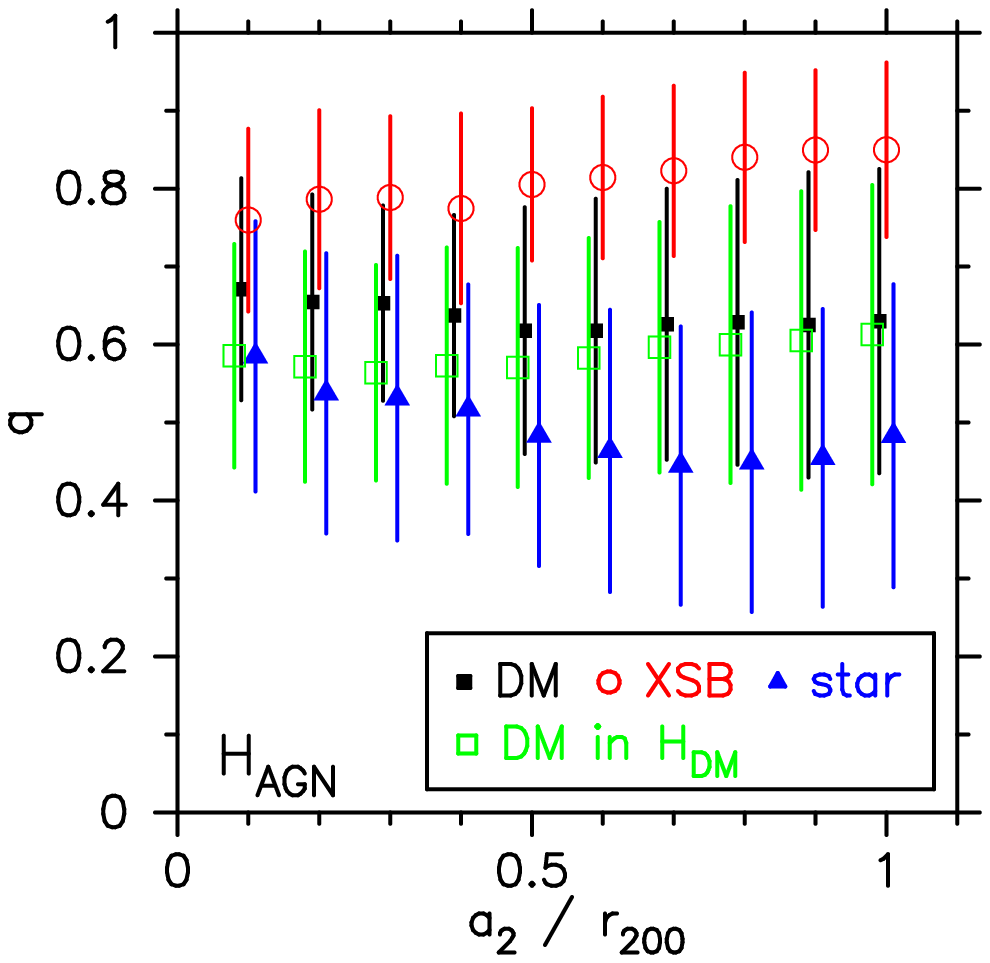}
\end{center}
\caption{Projected axis ratio $q$ of the X-ray surface brightness (red),
dark matter (black), and stars (blue) for the semi-major axis
$a_2/r_{200}=0.1$, 0.2, $\cdots$ 0.9, and 1.0, averaged over the $40\times 3$
clusters, for \HnoAGN~ (left) and \HAGN~ (right). The lines indicate the
standard deviation. To facilitate visualization, the results for dark
matter and stars are slightly shifted to the left and right,
respectively.}  \label{f5rq}
\end{figure}

{In Figure \ref{f5mq}} we find that the mass dependence of $q_{\rm XSB}$ is very weak
both in \HnoAGN~ and \HAGN, although their scatter is fairly large.  We
also find that $q_{\rm XSB}$ depends only weakly on the radius of galaxy
clusters.  This is illustrated in Figure \ref{f5rq}, where the mean and
standard deviation {of $q_{\rm XSB}$, $q_{\rm star}$, and $q_{\rm
DM}$} computed over all the simulated clusters are plotted against
$a_2/r_{200}$; \HnoAGN\, (left) and \HAGN\, (right).

The blue triangles in Figure \ref{f5rq} indicate that the density
distribution of stars is less spherical than that of dark matter for
both simulations. Also, the mean $q_{\rm star}$ is higher in \HnoAGN\,
than in \HAGN, as in the case of dark matter. The mean $q_{\rm star}$
tends to be larger toward the center for both simulations.  {This
may be partly interpreted as the effect of the cosmic web where more
galaxies at outer regions are accreted preferentially along the
filaments \citep{Aubert2004,Welker15}.}

Consider first \HnoAGN.  In the innermost region ($a_2=0.1r_{200}$),
dark matter distribution is more spherical than that of {XSB} and stars.
Since the gravitational potential there is dominated by stars (left
panel of Figure \ref{f5dtwo}), $q_{\rm XSB}$ is expected to be larger
than $q_{\rm star}$ if the HSE assumption holds.  Therefore Figure
\ref{f5rq} implies that gas in \HnoAGN~ is not in {HSE}
in the innermost region. This is most likely simply due to
the unrealistic over-cooling of gas in \HnoAGN, and indicates that the
baryon physics needs to be properly incorporated in simulations {in
order} to predict the non-sphericity of central regions of galaxy
clusters.

Next we examine \HAGN\, in detail.  As shown in the right panel of
Figure \ref{f5rq}, the mean value of $q_{\rm XSB}$ is roughly 0.8 and
very weakly increases from inner to outer regions. In contrast, both
$q_{\rm DM}$ and $q_{\rm star}$ decrease towards $a_2\approx 0.5
r_{200}$, and then become almost constant beyond the scale. For
reference, we plot $q_{\rm DM}$ for \HDM\, in green open squares.
{While the mean values of $q_{\rm DM}$ in \HDM\, and \HAGN\, are
almost the same for $a_2> 0.5 r_{200}$, $q_{\rm DM}$ in \HDM\, is} {significantly}
{smaller for $a_2< 0.5 r_{200}$.}  Again this illustrates that the
non-sphericity of galaxy clusters is significantly influenced, even up
to the half of their virial radius, by the baryon physics operating
around the more central regions. In other words, reliable predictions
for the non-sphericity of dark matter inside $0.5r_{200}$, approximately
corresponding to the mass scale of $M_{500}$, cannot be made with dark
matter only simulations, and require the hydrodynamical simulations with
well calibrated cooling and feedback effects.

\section{Statistical comparison of the projected axis ratio of 
X-ray surface brightness of simulated and observed galaxy clusters
\label{sec:comparison}}

We finally calculate the PDF of $q_{\rm XSB}$ of our simulated clusters,
and compare it with the data analyzed by \cite{Kawahara10}.  {His}
{sample of clusters is based on the
XMM-Newton cluster catalog compiled by \citet{Snowden08}. Their
selection of the sample is fairly empirical, but basically covers all
clusters that permit the measurement of the temperature
profile. \citet{Kawahara10} attempted the ellipse fit to all the 70
clusters in the catalog, and retained 61 clusters with the
signal-to-noise ratio exceeding unity at $a_2=0.1r_{200}$ in
constructing the PDF of the axis ratio. }
{Also, he considers the axis ratio from 61, 56, 39 and 13 clusters at
$a_2/r_{200}=0.1, 0.2, 0.3$ and $0.4$, respectively. This makes the total
number of available measurements 169.}

\begin{figure}[bthp]
\begin{center}
\FigureFile(80mm,80mm){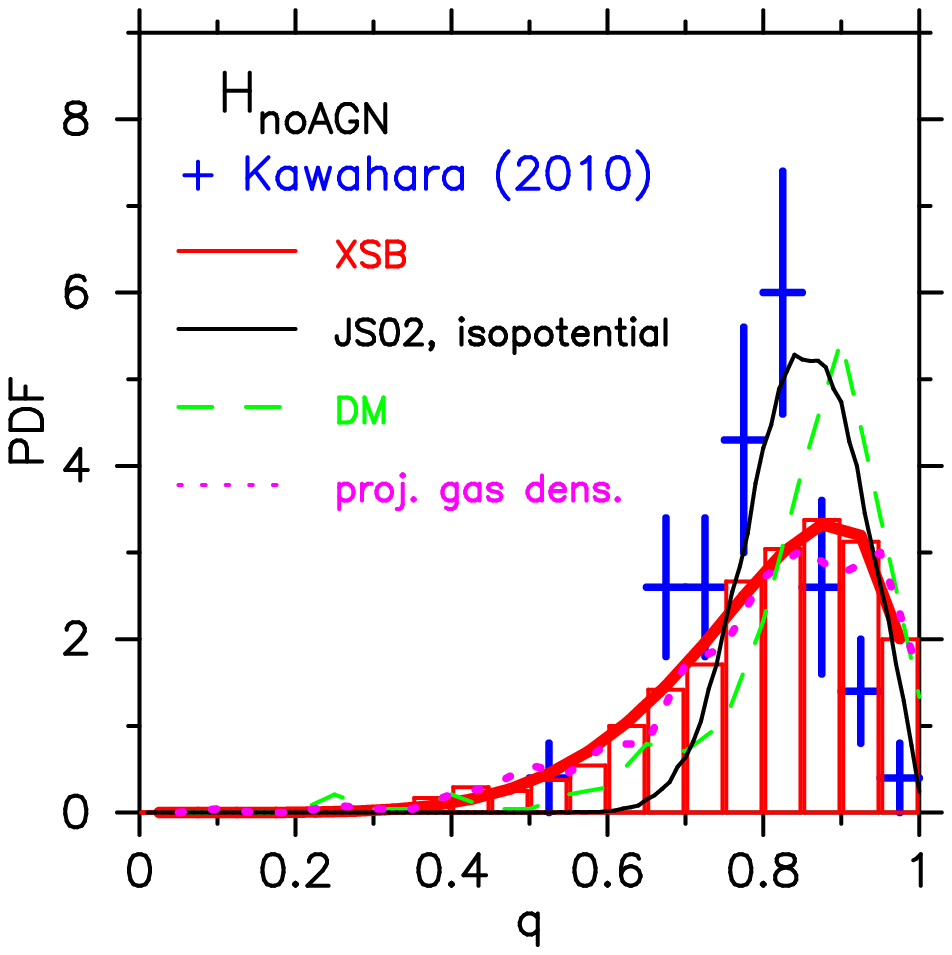}
\qquad
\FigureFile(80mm,80mm){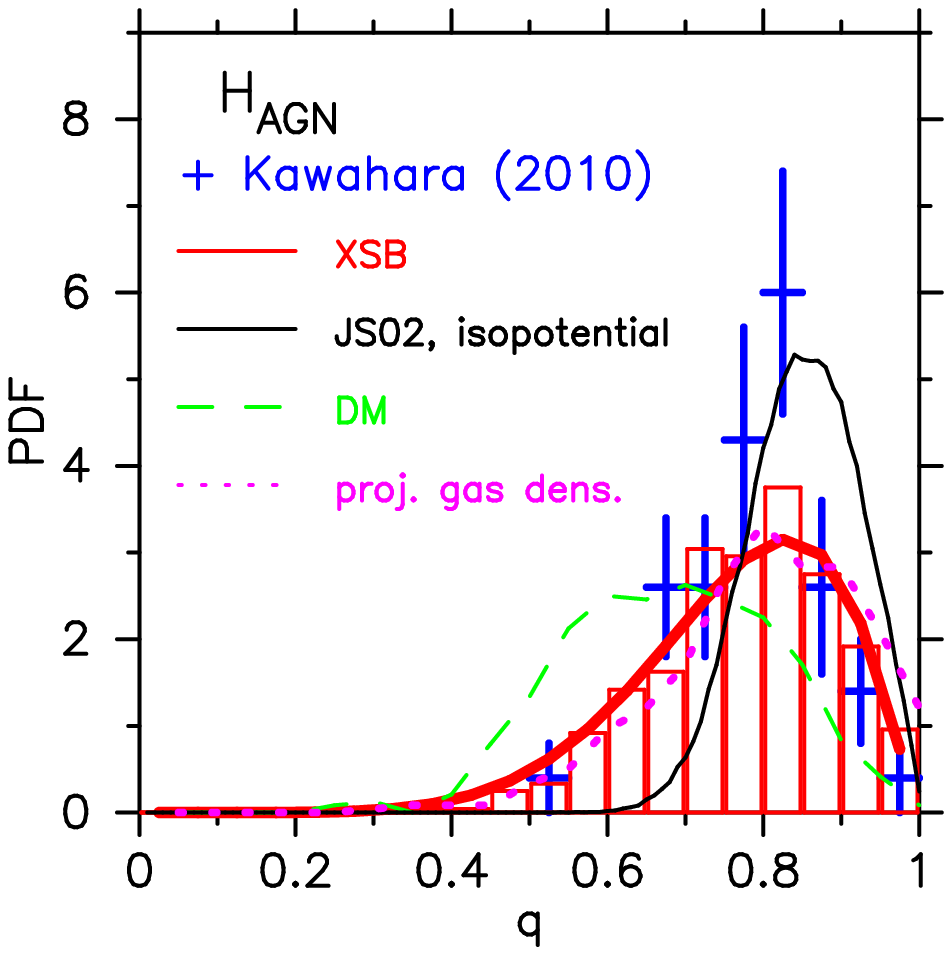}
\end{center}
\caption{PDFs of the projected axis ratio $q$; \HnoAGN~ (left) and \HAGN~
(right).  Results from numerical simulations are plotted for $S_X$ in
red histograms (fits to the $\beta$ distribution are plotted in red
thick lines), for dark matter in green dashed lines, and for gas in
magenta dotted lines.  The blue crosses with error-bars indicate the
observational data compiled by \cite{Kawahara10}.  For comparison, the
predictions based on the projected isopotential surfaces of self-similar
triaxial dark matter profiles \citep{Jing02,Lee03,Oguri03} are shown in
black solid lines.}  \label{f5distq}
\end{figure}

 Since we have seen that $q_{\rm XSB}$ is fairly insensitive to radius
and mass in the previous section, we combine the results for all the
four semi-major axis lengths $a_2/r_{200}=0.1$, 0.2, 0.3, and 0.4. The
range of $a_2/r_{200}$ is identical to that of \cite{Kawahara10}.  The
resulting PDF is plotted in Figure \ref{f5distq}. The number of the
cluster sample is 480 (40 halos $\times$ 3 lines-of-sight $\times$ 4
semi-major axis lengths) for each simulation, but strictly speaking they
are not necessarily independent.

In Figure \ref{f5distq}, the histogram corresponds to the PDF of $q_{\rm
XSB}$ for our simulated clusters.  For comparison, the PDF of $q_{\rm
DM}$ is over-plotted in green dashed line.  {Since the gas is
supposed to trace the isopotential, instead of isodensity, surface of
the total matter if} {HSE} {holds, their difference is
qualitatively well understood}.  {We also note that} the PDF of $q_{\rm
gas}$ for the projected gas density (magenta dotted lines) directly
calculated from the simulation is roughly the same as the PDF of
$q_{\rm XSB}$; while the projected gas density is proportional to
$n_{\rm gas}$ and $S_X$ is proportional to $n_{\rm gas}^2$, their shapes
turn out to be roughly the same.

As in \citet{Suto16b}, we find that the histogram of projected axis
ratio is reasonably well approximated by the beta distribution:
\begin{equation}
P(x; a,b)=\frac{x^{a-1}(1-x)^{b-1}}{B(a,b)},
\label{betad}
\end{equation}
where 
\begin{equation}
B(a,b)=\int_0^1x^{a-1}(1-x)^{b-1}dx
\end{equation}
is the beta function and $a$ and $b$ are the two fitting parameters.
Their best-fit values for the PDF of $q_{\rm XSB}$ are $(a,b)=(6.74,
1.70)$ for \HnoAGN\, and $(a,b)=(7.50, 2.31)$ for \HAGN, which are
over-plotted in thick red lines in Figure \ref{f5distq}.

We also plot {by the} black solid curve the PDF of $q$ for isopotential
surfaces based on the PDF of $q_{\rm DM}$ {modeled} by JS02 assuming
self-similar triaxial ellipsoids.  As was shown in \citet{Suto16b}, the
self-similar assumption of dark matter halos is not so accurate.
Therefore, the difference between the black solid curves and the green
dashed lines is explained by the combination of the break-down of the
self-similar assumption in JS02 and the baryon effects.  The above
result implies that the quantitative comparison with observations
requires the direct analysis of numerical simulations with appropriate
baryon physics as we performed here.

\cite{Kawahara10} derived the PDF of $q_{\rm XSB}$ from the observed 61
XMM-Newton clusters (plotted as blue crosses in Figure \ref{f5distq}),
and attempted a preliminary comparison with the PDF for isopotential
surfaces (black solid curve) based on the model of JS02.  We can perform
more quantitative comparison using the prediction based on {our}
hydrodynamical {simulation} (red histogram).

For that purpose, we use the cumulative PDF and apply the
Kolmogorov-Smirnov (KS) test in order to avoid the binning effect due to
the limited number of data. In general, {the KS test is defined 
as follows. We consider an empirical distribution function $F_n(x)$,
which is a cumulative distribution of $n$ samples of $x$. Then we consider}
a null hypothesis that ``$F_n(x)$ is {drawn} from $F(x)$'' {for 
any given cumulative distribution function $F(x)$.
Then} the KS statistic $D_n$ is defined by
\begin{equation}
D_n=\sup_x|F(x)-F_n(x)|.
\label{eq:KS}
\end{equation}
{It is known that} the quantity $\sqrt{n}D_n$ obeys the following {distribution} 
independently of $F(x)$ and $F_n(x)$:
\begin{equation}
p(\sqrt{n}D_n\le x)=1-2\sum_{i=1}^\infty(-1)^{i-1}e^{-2i^2x^2}.
\label{e5ksp}
\end{equation}
In addition, for a confidence level $\alpha$, $K_\alpha$ is defined by
$p(\sqrt{n}D_n\le K_\alpha)=1-\alpha$. Then, if $\sqrt{n}D_n>K_\alpha$,
the null hypothesis is rejected for the confidence level $\alpha$.

\begin{figure}[tbp]
\begin{center}
\FigureFile(80mm,80mm){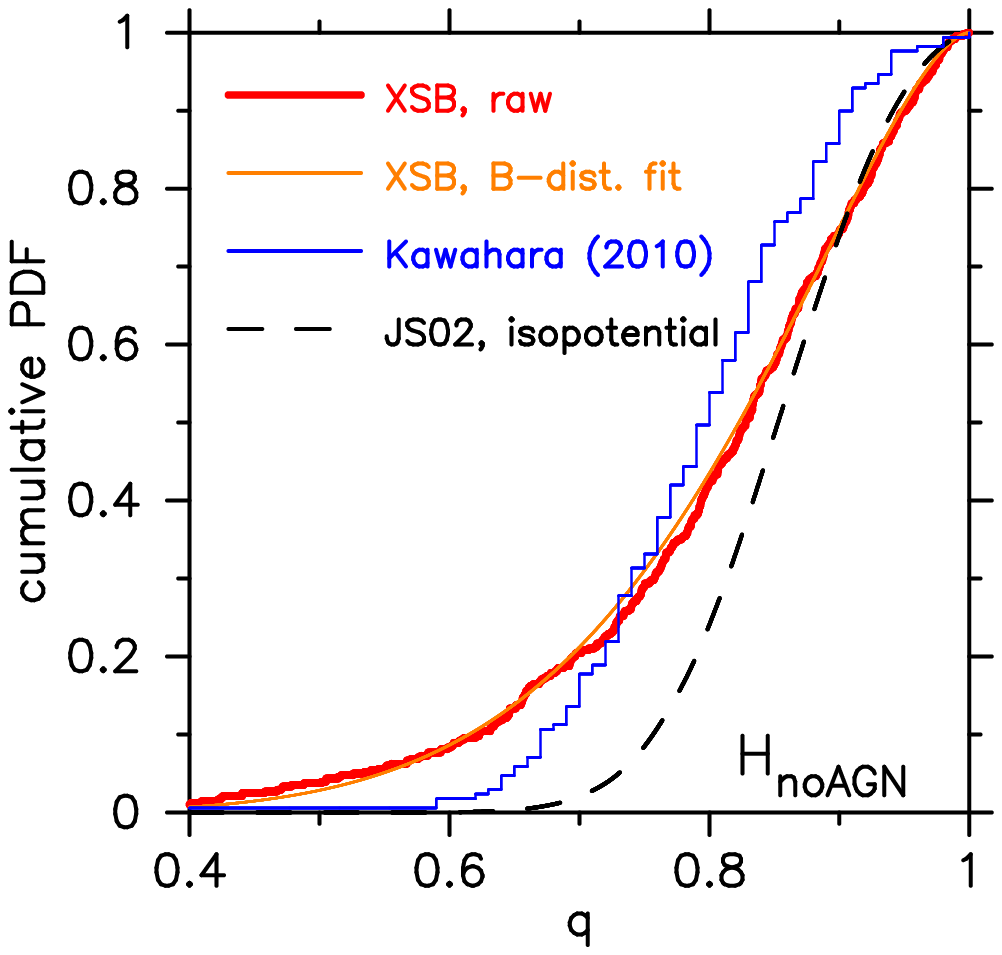}
\qquad
\FigureFile(80mm,80mm){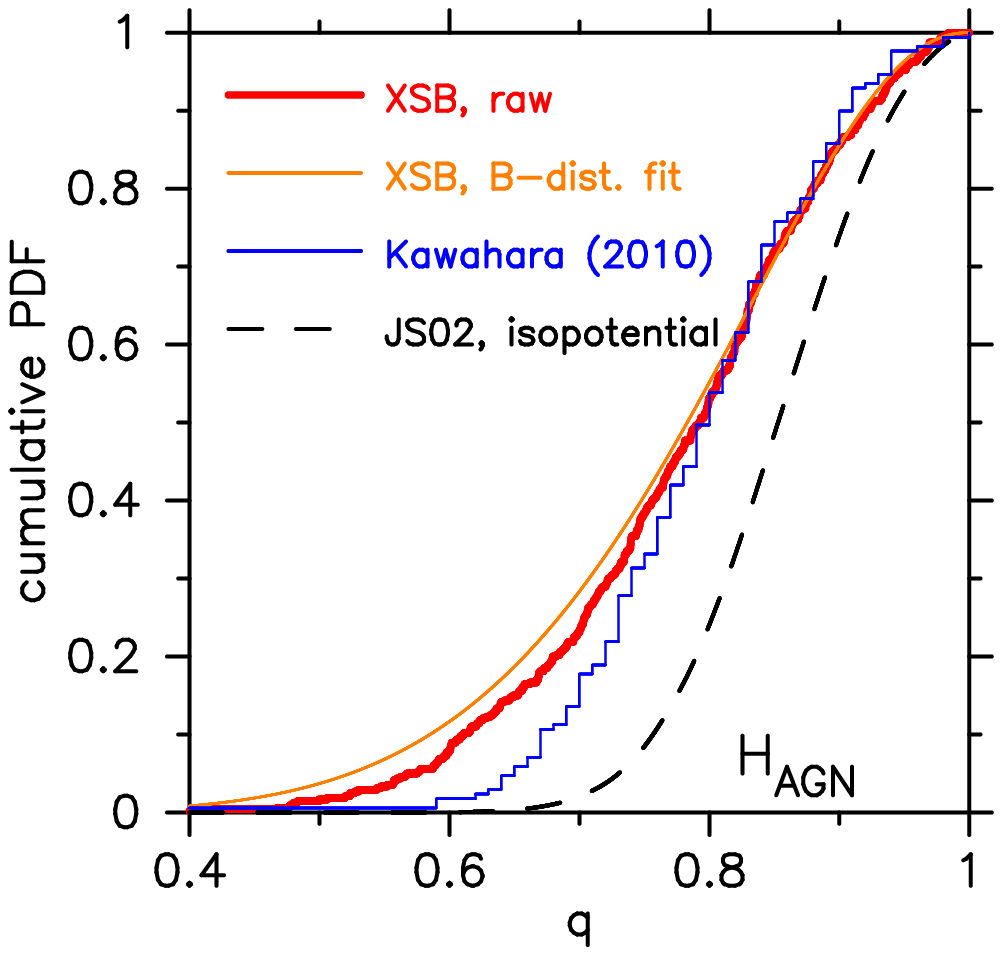}
\end{center}
\caption{Cumulative PDFs of $q$ of the X-ray surface brightness $S_X$;
\HnoAGN~ (left) and \HAGN~ (right). Results from numerical simulations are
plotted in thick red lines, and the corresponding fits on the basis of
$\beta$ distribution are plotted in thin orange lines against the
observational data in blue histograms compiled by \cite{Kawahara10}. For
comparison, the predictions based on the projected isopotential surfaces
of self-similar triaxial dark matter profiles
\citep{Jing02,Lee03,Oguri03} are shown in black dashed lines.}
\label{f5ks}
\end{figure}

First we {consider} the {observational} data analyzed by \cite{Kawahara10} and
the {prediction based on the isopotential surfaces of JS02 model
as $F_n(x)$ and $F(x)$}, respectively ($x=q$ and $n=169$). 
Figure \ref{f5ks} shows the cumulative PDFs of the observation data 
(thin blue solid) and the PDF of $q$ for
the isopotential surfaces (black dashed). Note that the {two
panels show the results based on \HnoAGN\,(left) and \HAGN\,(right), but these two lines
(blue and black) are independent of simulations and thus} the same for both panels. 
For these cumulative PDFs, we
obtain $\sqrt{n}D_n=3.89$, and the corresponding $\alpha$ is less than
$10^{-7}$. Hence the PDF of $q$ for the isopotential surfaces is
{highly} inconsistent with the observation data.

We next compare the observation data with our simulation results. {In
this case, both distributions are constructed from finite samples, thus we
consider a modification to the KS statistic (\ref{eq:KS}).}
For {two empirical distribution functions,}
$F_n(x)$ and $G_m(x)$ {respectively} with $n$ and $m$ samples, the
KS statistic {can be generalized as}
\begin{equation}
\bar{D}_{n,m}=\sup_x|F_n(x)-G_m(x)|.
\end{equation}
Then the quantity $\sqrt{nm/(n+m)}\bar{D}_{n,m}$ follows the same PDF as
Equation (\ref{e5ksp}).

{In this case,} we regard the {distribution} of $q$ for $S_X$ of our simulated clusters as
$G_m(x)$ ($x=q$ and $m=480$). The cumulative fraction for $q_{\rm XSB}$
is shown in thick red line in Figure \ref{f5ks} for \HnoAGN~ (left) and
\HAGN. For \HnoAGN, we obtain $\sqrt{nm/(n+m)}\bar{D}_{n,m}=2.04$, and
the corresponding $\alpha$ is less than $10^{-3}$. On the other hand,
for \HAGN, $\sqrt{nm/(n+m)}\bar{D}_{n,m}=1.24$, and the corresponding
$\alpha$ is 0.10. In other words, the probability that the observational
data {are drawn from the same distribution as} our simulation results 
is 10 \% for \HAGN, and less than 0.1 \% for \HnoAGN.

The agreement of our \HAGN~ result against the observation is not
perfect, but much better than that {of} the previous prediction based on
the various inaccurate {assumptions}
\citep{Lee03,Oguri03,Kawahara10,Suto16b}.  Thus we interpret this as
promising and encouraging even though it is premature to put any further
conclusion at this point. Instead we would like to note that there still
remain several issues that need to be carefully examined in both
observations and simulations. The statistical significance of the
present result is limited by the available number of observed clusters
with high-quality imaging data.  In particlar, the observational
data have much more weights at 0.1 and 0.2 $a_2/r_{200}$, while
simulation data have the same weight.  Also the mass ranges of the halos
do not exactly match between observations and simulations, although the
mass dependence is not so strong.  Such biases need to be taken into
account carefully.  The systematic uncertainty of the predictions is
dominated by the reliable implementation of baryon physics.  Those
issues will be improved with on-going and up-coming observational survey
projects, and we plan to carry out systematic numerical simulations in
order to address the reliability of the baryon physics {modeling}. We
hope to report the progress elsewhere in due course.

\section{Summary and Conclusion \label{sec:summary}}

 We have examined the non-sphericity of galaxy clusters using the
projected axis ratios of X-ray surface brightness, star, and dark matter
{distributions ($q_{\rm XSB}$, $q_{\rm star}$, and $q_{\rm DM}$)}.  We have
extracted 40 clusters of mass larger than $3\times10^{13}
M_\odot$ from cosmological hydrodynamic simulations that fully
incorporate baryon physics.

{In general, we find $q_{\rm XSB}>q_{\rm DM}$, which is
qualitatively understood by the fact that the gas roughly traces the
isopotential, instead of isodensity, surface of the total matter. Also
our simulation indicates that $q_{\rm DM}>q_{\rm star}$, but this should
be fairly sensitive to the baryon physics incorporated in the
simulation.  Indeed, the baryon physics, in particular the AGN feedback,
has a significant impact on the non-sphericity of galaxy clusters.} In
terms of the spherically averaged {\it dark matter} profile, the baryon
physics is important only in the region less than $\sim 10$\% of the
virial radius of each clusters. Nevertheless, its non-sphericity is
affected even up to the half of the virial radius. For instance, the
dark matter distribution becomes more spherical due to the baryon
{effect} relative to the dark matter only simulations. Also the trend of
increasing ellipticity of dark matter distribution against the radius is
opposite to that predicted from simulations neglecting baryon physics.
Only beyond the half of the virial radius, the non-sphericity of dark
matter distribution is not much affected by the baryon physics operating
in the central region. This should be kept in mind even when
weak-lensing analysis is used in order to measure the non-sphericity of
dark {matter} halos \citep{Oguri10,Suto16b}.

Then we have measured the projected axis ratio of X-ray surface
brightness, $q_{\rm XSB}$, for the 40 simulated clusters, and obtained
its probability density function. The latter is very different from the
previous prediction based on the projected isopotential surfaces of
self-similar triaxial dark matter profiles \citep{Jing02,Lee03,Oguri03},
indicating the importance of the direct estimate from the numerical
simulations. Therefore our resulting PDF significantly improves the
reliability of the prediction, which should be useful in future
observational confrontation.

Indeed while the previous prediction was not consistent with the
observational data compiled by \citet{Kawahara10}, our current improved
prediction based on the hydro-simulation with the AGN feedback exhibits
much better agreement. This is interesting and promising, but we admit
that it is not yet fully satisfactory: the statistics is severely
limited by the available number of high-quality clusters in {both}
observations and {in} simulations. Furthermore the parameter dependence of
the simulation, including cosmological parameters and empirical
parameters that control the baryon processes, needs to be examined
quantitatively.

Nevertheless our current study clearly indicates that the non-sphericity
of galaxy clusters will serve as a useful quantitative probe of
cosmology and cluster physics. The current methodology can be easily
applied to the cluster sample from the Sunyaev-Zel'dovich survey, e.g.,
\citet{Kitayama14,Kitayama16}, and also in principle to weak-lensing
\citep{Oguri03,Oguri10,Suto16b}. The joint analysis of X-ray surface
brightness, the Sunyaev-Zel'dovich effect, and weak-lensing in an
individual and/or statistical manner will {provide} an even more 
{powerful} approach to identify the nature of non-sphericity of galaxy clusters.
Future observational data and simulations will be able to test the cold
dark matter scenario and baryon physics simultaneously through the
non-sphericity of galaxy clusters in a complementary fashion to the
conventional statistics based on spherically averaged quantities.

\bigskip

\section*{Acknowledgements}

We thank Hajime Kawahara and Masamune Oguri for useful discussions.  The
present work is based on the Horizon simulation runs, which have been
performed using the HPC resources of CINES under the allocations
2013047012, 2014047012 and 2015047012 by GENCI.  The post-processed
analysis of the simulation was carried out on Cray XC30 at Center for
Computational Astrophysics of National Astronomical Observatory of
Japan, and also on the Horizon cluster at Institut d'Astrophysique de
Paris.  S.P. acknowledges support from the long-term invitation
fellowship by Japan Society for the Promotion of Science (JSPS).  This
work is supported partly by JSPS Core-to-Core Program "International
Network of Planetary Sciences", and by JSPS Grant-in-Aids for Scientific
Research No. 26-11473 (D.S.), No. 25400236 (T.K.), and No. 24340035
(Y.S.).



\end{document}